\documentclass [12pt]{article}
\usepackage{lscape}                                           %
\usepackage[centertags]{amsmath}
\usepackage{amsfonts} \usepackage{amssymb}
\usepackage{color}
\usepackage{graphicx}
\usepackage{lscape}
\usepackage{multirow}
\newcommand{\bu}{{\bar u}}

\newcommand{\cA}{{\cal A}}

\newcommand{\cC}{{\cal C}}

\newcommand{\cL}{{\cal L}}

\newcommand{\cN}{{\cal N}}

\newcommand{\uno}{\mathbb{I}}

\newcommand{\R}{\mathbb{R}}

\newcommand{\Hp}{\mathbb{H}}

\newcommand{\ii}{\imath}
\newcommand{\ee}{e}

\newcommand{\du}{\partial_u}

\newcommand{\oh}{\frac{1}{2}}
\newcommand{\noa}{}

\newcommand{\sqtap}{\sqrt{2 \alpha'}}
\newcommand{\tap}{ ({2 \alpha'}) }
\newcommand{\ap}{  \alpha' }

\newcommand{\mypiu}{-}
\newcommand{\mymeno}{+}

\newcommand{\subvertex}{color ordered vertex}
\newcommand{\subvertices}{color ordered vertices}
\newcommand{\Subvertices}{Color ordered vertices}

\newcommand{\amplitude}{amplitude}
\newcommand{\rA}{{A'}}

\newcommand{\hepsilon}{{\hat \epsilon}}
\newcommand{\he}{{\hepsilon}}
\newcommand{\hk}{{\hat k}}

\newcommand{\hs}{{\hat s}}
\newcommand{\hu}{{\hat u}}
\newcommand{\hx}{{\hat x}}
\newcommand{\hy}{{\hat y}}

\newcommand{\hX}{{\hat X}}

\newcommand{\qs}{{q}^*{}}
\newcommand{\hqs}{{\hat q}^*{}}

\newcommand{\cCz}{\cC_{0}}
\newcommand{\cNz}{\cN_{0}}

\newcommand{\COMMENTO}[1]{}
\newcommand{\COMMENTOBUH}[1]{}

\begin{document}
\title{
On the gauge chosen by the bosonic open string 
}

\author{
{Igor Pesando${}^1\,{}^2$}
\\
~\\
~\\
$^1$Dipartimento di Fisica, Universit\`a di Torino\\
and I.N.F.N. - sezione di Torino \\
Via P. Giuria 1, I-10125 Torino, Italy\\
~\\
$^2$LAPTh, Universit\'e de Savoie, CNRS\\
9, Chemin de Bellevue, 74941 Annecy le Vieux Cedex, France
\vspace{0.3cm}
\\{ipesando@to.infn.it}
}

\maketitle
\thispagestyle{empty}

\abstract{
String theory gives \(S\) matrix elements from which is not possible
to read any gauge information. 
Using factorization we go off shell in the simplest and most naive way
and we read which are the vertices suggested by string.
To compare with the associated Effective Field Theory it is natural to
 use \subvertices. The \(\alpha'=0\) \subvertices{} suggested by string theory are
 more efficient than the usual ones since the three gluon \subvertex{}
 has three terms instead of six and the four gluon one has one term
 instead of three.
They are written in the so called Gervais-Neveu gauge.
The full Effective Field Theory is in a generalization of the
Gervais-Neveu gauge with \(\alpha'\) corrections.
Moreover a field redefinition is required to be mapped
to the field used by string theory. 

We also give an intuitive way of understanding why string choose this
gauge in terms of the minimal number of couplings necessary to
reproduce the non abelian amplitudes starting from color ordered ones.
}
keywords: {String theory}
\\ preprint:{LAPTH-001/17}

\newpage


\section{Introduction and conclusions}

String theory is a good candidate for describing all the interactions
in Nature, gravity included.  This happens because in its spectrum
there are both massless spin 1 and spin 2 particles.  Nevertheless the
presence of these particles does not mean that they can be identified
with gauge bosons and the graviton.  This can only be established when
interactions are considered.  Therefore the study and the derivation
of effective field theory (EFT) actions (to be understood as 1PI
actions) from string theory is a very
well studied subject starting already at the beginning of 70s 
(\cite{Scherk:1971xy}-\cite{Yoneya:1974jg})
and improved in the 80s (see for example \cite{Tseytlin:1986ti}) 
 but we want to approach
it from a different point of view.

Usually the aim is to determine the gauge invariant effective field
theory.

Our main focus is slightly different since we are not mainly
interested in the derivation of EFT action for
gluons but we want to explore in agnostic way which is the 
gauge fixed EFT suggested by string theory and the connection between
the fields used by string theory and the canonical 
ones usually used in defining EFT.
Essentially we will derive and extend the gauge proposed in
\cite{Gervais:1972tr}.
Our approach differs from the one used in \cite{Coletti:2003ai} since
we use plain old string theory and we are interested in finding the
gauge fixing suggested by it.
It differs also from \cite{Magnea:2015fsa} since we try to read the
gauge and the fields suggested by string theory rather than try to
verify that the gauge suggested in \cite{Gervais:1972tr} (or more
precisely an extension to the background field method in the case of
\cite{Magnea:2015fsa} as first proposed in \cite{Bern:1991an}) works.

\begin{figure}[hbt]
\begin{center}
\def\svgwidth{350px}
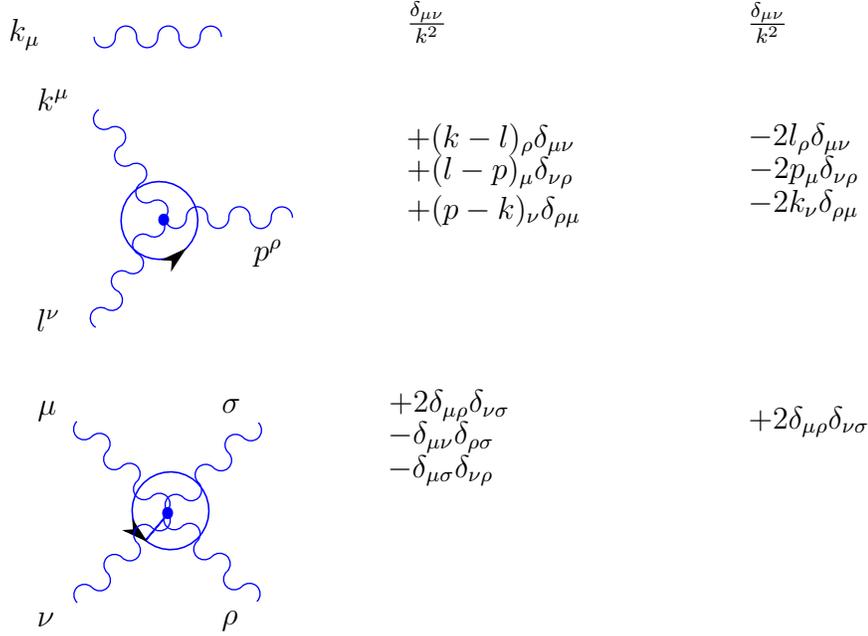
\end{center}
\vskip -0.5cm
\caption{
The usual Euclidean \subvertices{} vs the ones suggested by string
theory.
Each tree diagram must be then multiplied by 
$g^{N-2} \frac{1}{\kappa} Tr(T_{a_1}\dots T_{a_N})$ with $g$ the
Yang-Mills coupling constant, 
$N$ the number of legs
and $T_a$ the unitary algebra matrix normalized as in appendix \ref{sec:YMconv}.
}
\label{fig:Color_Ordered_Feynman_Rules}
\end{figure}

Since all choices made by string theory are clever it is worth trying
to read in the most direct way what it suggests. 
Actually it turns out that the suggested \subvertices{} by string theory
\cite{Gervais:1972tr} are
 more efficient than the usual ones since the three gluon \subvertex{}
 has three terms instead of the usual six 
and the four gluon one has one term instead of the usual three as
shown in figure \ref{fig:Color_Ordered_Feynman_Rules}. 
The reason is that the usual \subvertices{} are obtained starting from
the Feynman rules in Feynman gauge and then mimicking string by
performing a color decomposition (see
\cite{Mangano:1990by,Dixon:1996wi} and references therein), here we
adopt a more radical point of view and we try to mimick string in all.

Our starting point is to notice that while computing the EFT 
one is actually using  a gauge fixed EFT action.
The gauge fixing is necessary in order to have a well defined
propagator and a well defined propagator is needed in order to compute
the \(S\) matrix elements which are then compared with the ones from
string theory.
This happens because the EFT is a 1-PI action and the $S$ matrix
elements are computed from truncated on shell Green functions.

\begin{figure}[hbt]
\begin{center}
\def\svgwidth{350px}
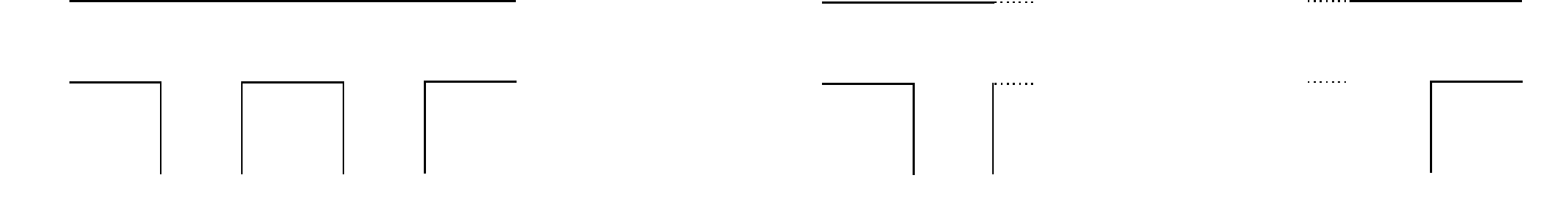
\end{center}
\vskip -0.5cm
\caption{
The 4 state  string \(S\) matrix is given by the sum of the product of two
3 state string amplitudes and a propagator 
where the intermediate state \(i^{*}\) is not required to be physical. 
}
\label{fig:N4_factorization}
\end{figure}

\begin{figure}[hbt]
\begin{center}
\def\svgwidth{450px}
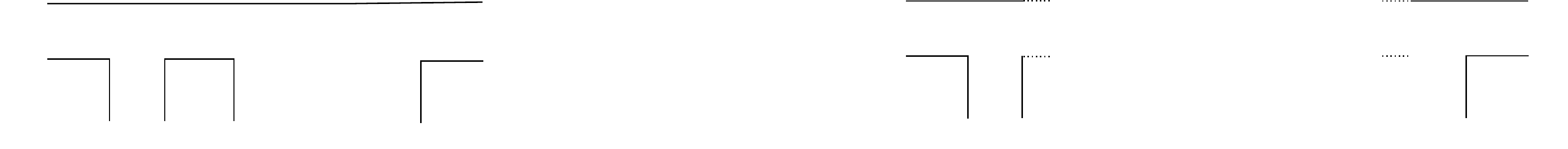
\end{center}
\vskip -0.5cm
\caption{
The 5 state string \(S\) matrix is given by the sum of the product of three
3 state string amplitudes and two propagators 
where the intermediate \(3\) state string amplitude has two states
\(i^{*}, j^{*}\) which are not required to be physical. 
}
\label{fig:N5_factorization}
\end{figure}

The $S$ matrix elements of gauge invariant operators are obviously
independent on the gauge fixing but the intermediate steps are not.
So one could wonder how it is possible to extract any information on
gauge fixing and fields comparing \(S\) matrices.
In fact it is not possible.
Nevertheless factorization of string amplitudes allows to have a glimpse
on how string theory fixes the gauge  since it yields amplitudes with
off shell and unphysical states (see \cite{Cuomo:2000de} for previous
work on how to extend off shell the string amplitudes).
For example figure \ref{fig:N4_factorization} shows how it is possible
to obtain a \(3\) state amplitude with one possibly unphysical state from a
factorization of a \(4\) state physical one. 
In the same spirit it is possible to start with a $5$ state amplitude
and get an amplitude with one physical state and two possibly not
physical ones as shown in figure \ref{fig:N5_factorization}.

Using these unphysical amplitudes  we can try to understand which
gauge fixing is suggested by string theory in the EFT
computations. 
To find which gauge is chosen by string we have actually to introduce
some other requirements.
The reason is the following.
The amplitude with two possibly non physical states is figure
\ref{fig:N5_factorization} is not the full \(3\) point truncated Green
function, i.e. the $3$ vertex.
It is only a part of it since Green functions are totally symmetric on
the external legs and the amplitude we get is not.
This means that either we compute the full $3$ vertex or we
compare with a \subvertex.

The first approach is not readily available 
since the off shell \(3\) point string partial amplitude treats 
in asymmetric way the off shell gluons\footnote{
This issue can be probably avoided using the twisted propagator at the
price of having a non canonical propagator (see
\cite{Alessandrini:1971fx} and references therein).
The issue is under investigation.}. 
This means that  if we want to construct a \(3\) vertex, that is
required to be totally symmetric in the exchange of gluons, we should
sum over all the permutations of the external states. This would
require to reexam the way we are used to do string computations and it
would lead too far away.

We are therefore left with to the latter approach,
also for ease of computation.
In doing this partial identification 
then we introduce an element of arbitrarily.
Since each on shell string diagram is cyclically invariant it is the natural to
compare with a cyclically invariant \subvertex.
It turns out that the $3$ string truncated Green function 
we are dealing with is not
cyclically invariant but it is up to gauge conditions.
It follows then that we cannot identify the $3$ state string truncated
Green function   with the cyclically invariant \subvertex{} but there is a left
over, see eq. (\ref{V3-gluons-symmetric+diff}).
This means  that we cannot exactly match the naive off shell string
amplitudes  with an EFTbut we can try to mimic them as close as possible.
We are therefore left with the choice of how to choose the left over. 
Then the result on the gauge then depends 
on the assumption on what it means to mimic as close as
possible the string truncated Green functions with the EFT vertices.
Obviously we are not obliged to use the suggested gauge and use
whichever gauge we want but trying to mimic as close as possible the
string can give useful ideas.
Our way of defining as close as possible is to try to minimize the
number of left over terms in the \(3\) state vertex and then check
that this implies that the number of terms in \(4\) point
\subvertex{}, i.e. the contact terms, is also minimized.
This is what done in this paper.

Using this approach we find that the gauge chosen is an \(\alpha'\)
corrected  version of the Gervais-Neveu gauge \cite{Gervais:1972tr} 
and that the field chosen by string theory to describe the gluon
is not the gauge field used naturally in EFT 
but it is connected to it by a field redefinition.
This kind of field redefinition is natural and expected in string
field theory but it is a kind of surprise in the plain old string
theory.
Since at the end we are comparing partial color ordered 
\(S\) matrix elements 
we can also  use a gauge fixed EFT expressed using 
the usual gauge field and the usual Feynman
gauge  at the price of having
a bigger difference between the vertices suggested by the string and
the ones computed from EFT.

It would also be interesting to consider the \subvertices{} suggested
by string theory in a magnetic background using the tecniques
developed in \cite{Pesando:2014owa,miei_prima}
and compare with the ones
used in \cite{Magnea:2015fsa}. It is very likely that the string
suggestion is of a non-commutative nature.
Also considering the superstring could be interesting in order to see
whether a field redefinition is necessary.

The rest of this article is organized as follows.
In section \ref{sect:main_idea} we describe in more details the idea
on how to read the vertices and \subvertices{} from string theory and
we compare with the usual approach in determining the EFT.
We introduce the \subvertices{} in a slightly different way as usual
(see \cite{Mangano:1990by,Dixon:1996wi} and references therein)
since they are introduced as a tool to mimic string diagrams as close
as possible.
In section \ref{sec:3points} we perform the actual computation of the
\(3\) \subvertex{}. We discuss how it compares to the most general
\(3\) vector Lagrangian and 
the field redefinition which is needed to map the string
field to the usual one used in EFT.
We also discuss the string \subvertex{} as result of the
minimal information which is needed to reconstruct the gauge invariant EFT.
Finally in section \ref{sec:4points} we recover the \(4\) point
\subvertex{} up to two derivatives and we show that choice performed for
the \(3\) vertex is the one which minimize the number of terms in this
\(4\) point vertex.

\COMMENTO{
We are also interested in computing in an efficient way effective
vertices with many ($\geq 5$) legs.
The usual way is to compute the full string amplitudes and the
subtract the reducible Feynman diagrams which involve effective
vertices with less legs.
We show that this can be done in an efficient way for vertices with
less derivatives by carefully factorizing the amplitudes on the states
which produce fewer derivative.
It turns out that these states are all  auxiliary  (spurious?)
}

\section{The basic idea}
\label{sect:main_idea}
In this section we would like to summarize some  well known facts and
then explain in more detail the basic idea behind this paper.
The first point to quickly review is  how factorization works in the
simplest setting and allows to extract string amplitudes where some
states are not required to be physical.
Then we review the connection between Lagrangian interactions and
Feynman vertices and we discuss the \subvertices{}  
(see \cite{Mangano:1990by,Dixon:1996wi} for a different way of
introducing them)
which are then used in the rest of the paper for extracting the gauge
fixing.
Finally we exemplify the approach with the simplest computation,
i.e. the derivation of the propagator or that is the same the kinetic term.

\subsection{Simple factorization}

In the old days of string theory the tree 
amplitude of \(N\) open string physical
states \(\phi_{i}\) (\(i=1,\dots N\)) was computed as
(see appendix  \ref{app:conventions} for conventions)
\begin{equation}
\label{N-old-old-ampl}
A(\phi_{1},\dots \phi_{N})
=
\langle\langle \phi_{1}| 
V(1; \phi_{2})\frac{1}{L_{0}^{(X)}-1}
V(1; \phi_{3})\dots
\frac{1}{L_{0}^{(X)}-1} V(1; \phi_{N-1})
|\phi_{N}\rangle
.
\end{equation}
This amplitude is cyclically symmetric, i.e.
\(A(\phi_{1},\dots \phi_{N})
=A(\phi_{N},\dots \phi_{1})
\).
In the previous expression 
\(V(x; \phi)\) is the vertex operator associated to the physical
state \(\phi\) of conformal dimension 1,
\(|\phi\rangle = V(x=0, \phi) |0\rangle_{SL(2,\R)}\).
This expression roughly corresponds to a truncated Feynman diagram
associated with a cubic theory and propagator
\({1}/{(L_{0}^{(X)}-1)}\).
Truncated diagram because the states are on shell and because of this
there is not propagator immediately after (before) the bra(ket) state.

The previous expression gives part of the \(S\) matrix and
the full \(S\) matrix is obtained by summing over all non cyclically
inequivalent permutations after having multiplied the previous
expression for the Chan Paton contribution and having given a color
\(a\) to all the physical states \(\phi \rightarrow \phi_{a}\),
explicitly
\begin{align}
S(\phi_{1, a_{1}},\dots \phi_{N, a_{N}})
=&
\ii ~\cA(\phi_{1, a_{1}},\dots \phi_{N, a_{N}})
,
\end{align}
where 
\(\cA\) 
is the connected truncated Green function
\begin{align}
\cA(\phi_{1, a_{1}},\dots \phi_{N, a_{N}})
=&\frac{ \alpha'^{N-3}}{\kappa} \cCz \cNz^{N}
\sum_{\mbox{non cyclical perm.s \(\sigma\)}}
A(\phi_{\sigma(1),  a_{\sigma(1)}},\dots \phi_{\sigma(N), a_{\sigma(N)}})
~tr(T_{a_{\sigma(1)} \dots } T_{a_{\sigma(N)}})
,
\nonumber\\
\end{align}
where the factor \(\alpha'^{N-3}\) can be reabsorbed into the definition
of the tree amplitude normalization \(\cCz\)
\cite{DiVecchia:1996uq} and the vertex
normalization \(\cNz\) but we prefer to make it clear since it makes 
the propagator canonical%
\footnote{
Explicitly we have with respect to \cite{DiVecchia:1996uq}
\(\cCz^{\mbox{here}}= \cCz^{\mbox{there}} \kappa \alpha'^{3} \) and
\(\cNz^{\mbox{here}}= \cNz^{\mbox{there}} /\alpha' \)
when we consider the different trace normalizations
\(tr^{\mbox{there}}(T_{a}T_{b}) = \oh \delta_{a b}\) while we use the
normalization given in eq.
(\ref{eq:tr_norm}) which implies \(\kappa=\oh\). 
}.


The previous amplitude (\ref{N-old-old-ampl}) can be recast in a more 
modern form by writing the propagator in an integral form
\({1}/{(L_{0}^{(X)}-1)} = \int_{0}^{1}dy~y^{L^{(X)}_{0}-2}\)
and then moving all the terms involving \(L^{(X)}_{0}\) to the right
and changing integration variables to get a correlator integrated over
the moduli space as
\begin{equation}
\label{N-old-ampl}
A(\phi_{1},\dots \phi_{N})
=
\int_{0}^{1} d x_{3} \int_{0}^{x_{3}} d x_{4}\dots \int_{0}^{x_{N-2}}d x_{N-1}~
\langle\langle \phi_{1}| 
V(1; \phi_{2})
V(x_{3}; \phi_{3})\dots
 V(x_{N-1}; \phi_{N-1})
|\phi_{N}\rangle
.
\end{equation}

For our purposes we need only the 3 point amplitude with two non physical
states and one physical which can be readily obtained by factorizing
the \(N=5\) amplitude in the old form
\begin{equation}
A(\phi_{1},\dots \phi_{5})
=
\langle\langle \phi_{1}| 
V(1; \phi_{2})\frac{1}{L_{0}^{(X)}-1}
V(1; \phi_{3})
\frac{1}{L_{0}^{(X)}-1} V(1; \phi_{4})
|\phi_{4}\rangle
.
\end{equation}
We can now insert four times the partition of unity
\begin{align}
\uno
=&
\int \frac{d^{D}\hk}{(2\pi)^{D}}
\Bigl[ | \hk \rangle~\langle\langle \hk |
+ \alpha^{\mu}_{-1}| \hk \rangle~\langle\langle \hk |\alpha^{\mu}_{1}
+ { \alpha^{\mu}_{-2}\over \sqrt{2} }| \hk \rangle
~\langle\langle \hk |{\alpha^{\mu}_{2} \over \sqrt{2} }
+ {\alpha^{\mu}_{-1}\alpha^{\nu}_{-1} \over \sqrt{2!} }| \hk \rangle
~\langle\langle \hk |{\alpha^{\mu}_{1}\alpha^{\nu}_{1} \over \sqrt{2!} }
+\dots
\Bigr]
\nonumber\\
=&
\sum_{\alpha} |\alpha\rangle ~\langle\langle \alpha |
,
\end{align}
where $\hk$ is the dimensionless momentum and
\( |\alpha\rangle \) is a generic basis element of the string Fock
space which is eigenstate of \(L_{0}^{(X)}\) with eigenvalue
\(l_{0}(\alpha)\).
These states are normalized as
 \( \langle\langle \beta | \alpha\rangle = \delta_{\alpha,\beta}\).
We then immediately get the mathematical expression corresponding to
figure \ref{fig:N5_factorization}
\begin{equation}
A(\phi_{1},\dots \phi_{5})
=
\sum_{\alpha, \beta}
\langle \phi_{1}| 
V(1; \phi_{2}) |\alpha\rangle
\frac{1}{l_{0}(\alpha)-1}
\langle\langle \alpha | V(1; \phi_{3})| \beta \rangle 
\frac{1}{l_{0}(\beta)-1} 
\langle\langle \beta | V(1; \phi_{4}) |\phi_{4}\rangle
.
\end{equation}
In this expression the sub-amplitude with two states which are not
necessarily physical is  
\( \langle\langle \alpha | V(1; \phi_{3})| \beta \rangle  \)  
and corresponds to the  part of the figure \ref{fig:N5_factorization}.
with dotted lines.
Notice however that this amplitude is not cyclically symmetric as the
corresponding amplitude with physical states (it is however actually
sufficient to have off shell but transverse states to get cyclicity).

\begin{figure}[hbt]
\begin{center}
\def\svgwidth{450px}
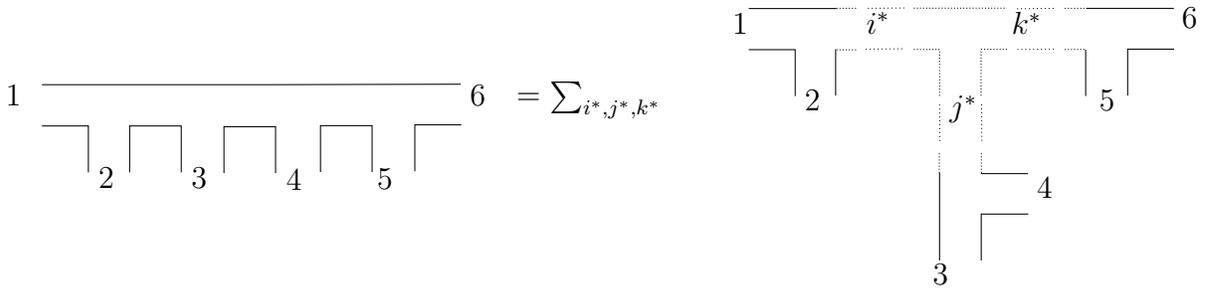
\end{center}
\vskip -0.5cm
\caption{
The 6 states string \(S\) matrix is given by the sum of the product of four
3 states string amplitudes and three propagators where one of the 3
states vertices involves 
three \(i^{*}, j^{*}, k^{*}\) states which are not required to be physical. 
}
\label{fig:N6_factorization}
\end{figure}

Even more generally 
starting from a \(6\) state amplitude is possible to find a \(3\)
state amplitude where all states are possible unphysical as shown in
figure \ref{fig:N6_factorization} and first derived in the seminal
paper \cite{SDS}.
In the rest of the paper we are not going to use this more general
vertex and therefore we do not write its expression.

\begin{figure}[hbt]
\begin{center}
\def\svgwidth{250px}
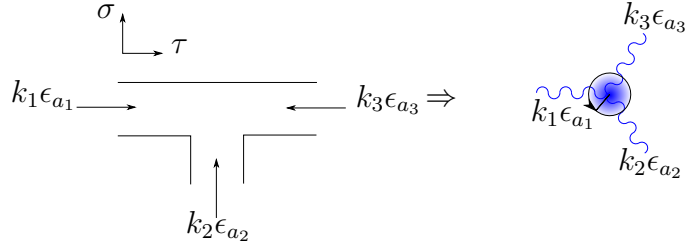
\end{center}
\vskip -0.5cm
\caption{
The intuitive reason why we label the states in a counterclockwise fashion.
}
\label{fig:why_CCw}
\end{figure}

In the following we depict the string amplitudes  mostly as
interactions on a disc.
On a disc the states are labeled counterclockwise because this is the natural
way of labeling starting from the intuitive strip picture as shown in
figure \ref{fig:why_CCw} for the three gluon amplitude.
 
\begin{figure}[hbt]
\begin{center}
\def\svgwidth{250px}
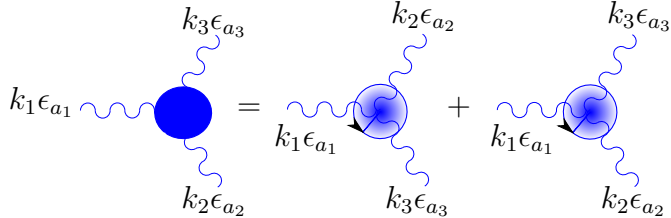
\end{center}
\vskip -0.5cm
\caption{
The 3 gluon string \(S\) matrix is given by the sum of the two cyclically
inequivalent orderings.
}
\label{fig:V3_string}
\end{figure}

In the case of 3 gluons the string \(S\) matrix element can then be
depicted as in figure \ref{fig:V3_string}.

\COMMENTOBUH{
Ambiguity due to vanishing of 3 point amplitudes when one state is
BRST exact??
}

\subsection{Usual way of computing the EFT}

To compute the gauge invariant EFT we proceed order by order in the
number of fields $A^N$,  in power of derivatives $\partial^n$ 
and in the YM coupling constant $g^k$.
The Lagrangian of order $N$ can be written schematically as
$
{\cL}_{[N]}= \sqtap^{-D}$ %
$\times[ \sqtap^{2-\oh D} g]^k$ %
$\times [ \sqtap \partial]^m$ %
$\times[\sqtap^{\oh D -1} A]^N$.
Taking in consideration that $g$s originates efficaciously from cubic
vertices in string we have the usual relations $ 3 k = 2 I + N$ and 
$I=L+k-1$ where $I$ is the number of internal lines.
We can then write
$
{\cL}_{[N]}= \sqtap^{N+m-L(D-4)}  g^{N-2+2 L} \partial^m A^N
$
where $L$ is the number of loops.
Since the Lagrangian is a scalar and we want it to be expressed using
gauge invariant field strength (we do not consider Chern-Simons
theories) we need an even number of Lorentz indeces and we get finally
$
{\cL}_{[N]}= \sqtap^{N+n-4 -L(D-4)}  g^{N-2+2 L} \partial^n F^N
$.

In the following we are interested in the tree EFT, i.e. $L=0$.
If we are also only interested  
to up $\tap^n$  then only a finite number of terms are needed
since $N\le \oh n +2$.

The usual procedure for computing the EFT is roughly as follows.
Suppose we have computed the EFT to order \(N-1\) in the fields and
order $\tap^{n}$.
In order to do so we have fixed a gauge 
since in order to compute the \(S\) matrix elements with $k$ particles
we need  the $k$ point Green functions and they are obtained from 1PI
vertices also by joining some of them with inverse propagators.
To compute the next order involving \(N\) fields  then
\cite{Tseytlin:1986ti}:
\begin{itemize}
\item
write down the most general gauge invariant Lagrangian with at
least \(N\) fields;
\item
check that all terms are independent;
\item
consider all the field redefinitions 
with at most \(N\) fields which do not change the $S$ matrix (see 
\cite{Tseytlin:1986ti}
for a discussion for the open string theory) 
and how these field redefinitions change the
coefficients of the independent terms of the Lagrangian;
\item 
determine which combinations of the coefficients are left invariant by
field redefinitions;
\item 
compute a number of \(S\) matrix elements with \(N\) fields sufficient
to determine the independent combinations of the coefficients
\item
compare the previous \(S\) matrix elements with the corresponding ones
from string theory in order to fix explicitly the independent combinations.
\end{itemize}

Consider the Euclidean Lagrangian up to
  $N=4$  and $\tap^2$ orders we have order by
order in $N${}\footnote{
Note that due to Bianchi identity we have \cite{Tseytlin:1986ti}
$
tr( D_\rho F_{\mu \nu}~D_\rho F_{\mu \nu})
\equiv 2 tr( D_\rho F_{\rho \mu}~D_\sigma F_{\sigma \mu}
- 2 F_{\rho \sigma} ~F_{\rho\lambda} ~F_{\sigma\lambda}
)
$
up to total derivatives.
}
\begin{align}
S_{E\,[2]} = \int d^D x\,
\frac{1}{{\kappa}}
tr \Bigl[
&\frac{1}{4} F_{\mu \nu} F_{\mu \nu}
+ \tap \Bigl(
+ v_{[2] 1} D_\rho F_{\rho \mu}~D_\sigma F_{\sigma \mu}
\Bigr)
+ \tap^2 \Bigl(
v_{[2] 2} D_\rho D_\sigma F_{\mu \nu}~D_\rho D_\sigma F_{\mu \nu}
\Bigr)
\Bigr]
\end{align}

\begin{align}
S_{E\,[3]} = \int d^D x\,
\frac{1}{{\kappa}}
tr \Bigl[
&\tap v_{[3] 0} F_{\mu \nu} F_{\nu \lambda} F_{\lambda \mu}
\nonumber\\
+ \tap^2 
&\Bigl(
v_{[3] 1}  F_{\mu \nu} ~D_\rho F_{\mu \nu} ~D_\sigma F_{\sigma \rho}
+ v_{[3] 2} F_{\mu \nu} ~D_\sigma F_{\sigma \rho} ~D_\rho F_{\mu \nu}
+v_{[3] 1}  F_{\mu \nu} ~D_\rho F_{\rho \mu} ~D_\sigma F_{\sigma \nu}
\Bigr)
\Bigr]
\end{align}

\begin{align}
S_{E\,[4]} = \int d^D x\,
\frac{1}{{\kappa}}
tr \Bigl[ \tap^2 \Bigl(
& 
v_{[4] 0} F_{\mu \nu} F_{\nu \lambda} F_{\lambda \kappa} F_{\kappa \mu}
+ v_{[4] 1} F_{\mu \rho} F_{\mu \sigma} F_{\lambda \rho} F_{\lambda \sigma}
+ v_{[4] 2} F_{\mu \nu} F_{\mu \nu} F_{\rho \sigma} F_{\rho \sigma}
\Bigr)
\Bigr]
\end{align}
As usual there is an ambiguity on how to write the derivative terms
since
$[ D_\mu, D_\nu]\sim F_{\mu \nu}$.
Then we can also write the gauge fixing Lagrangian
\begin{align}
\label{eq:S_g.f.}
S_{E\,gf} = \int d^D x\,
&
\frac{\xi}{{\kappa}}
tr 
\Bigl(
\partial_\mu A_\mu
+ \tap \frac{g_0}{\xi} \partial^2 \partial_\mu A_\mu
+ \tap^{2-D/2} \frac{g_1}{\xi} A_\mu A_\mu
\nonumber\\
+&
\tap^{D/2}
\Bigl[
 \frac{g_2}{\xi} \partial_\mu A_\mu \partial_\nu A_\nu
+ \frac{g_3}{\xi} \partial_\nu A_\mu \partial_\mu A_\nu
+ \frac{g_4}{\xi} \partial_\mu A_\nu \partial_\mu A_\nu
+ \frac{g_5}{\xi} \partial^2 A_\mu A_\mu
+ \frac{g_6}{\xi} A_\mu \partial^2 A_\mu
\Bigr]
+\dots \Bigr)^2
.
\end{align}

Finally we can consider the field redefinitions.
We can consider field redefinitions which do no change the gauge
transformations like
\begin{equation}
A_\mu= A'_\mu + r D'_\rho F'_{\rho \mu} + \dots
\end{equation}
or we can consider field redefinitions which do change the gauge
transformations. 
If we are willing to change the gauge transformation then the only 
constraints are that all terms belong to the original algebra and that
they do no change the $S$ matrix elements.
We will see that we need such more drastic field redefinitions in
order to accomplish our program.
They are like
\begin{align}
\label{eq:field_red}
A_\mu=
& \rA_\mu 
+\tap \left(
r_1 \partial_\mu \partial_\rho\rA_\rho
+
r_2 \partial_\rho \partial_\rho\rA_\mu
\right)
\nonumber\\
&+
\tap^{D/2}
\left(
r_3 [\rA_\mu,  \partial_\rho\rA_\rho]
+r_4 [\rA_\rho,  \partial_\rho\rA_\mu]
+r_5 [\rA_\rho,  \partial_\mu\rA_\rho]
\right)
+\dots
\end{align}

The usual approach would then continue by finding the coefficients
$v$s which are left unchanged by field redefinitions and then fix them
by comparing the $S$ matrix elements.
This comparison is obviously independent on the gauge fixing.

\subsection{The approach and the propagator}
Differently from the usual approach the idea we want to implement is  first
to write blindly the EFT vertices mimicking the amplitudes with off
shell/unphysical states computed from the string.
Then to map these vertices to a gauge fixed EFT and determine
the necessary field redefinitions at the same time.

To see how this work let us consider the propagator, i.e. the case $N=2$.
From the previous discussion we know that the propagator is given
 by
\begin{align}
\langle\langle \hk_1, \mu_1|
\frac{\alpha'}{L_{0}^{(X)}-1}
|\hk_2, \mu_2\rangle
=
\frac{\delta^{\mu_1\mu_2}}{ \hk_1^2 } 
\delta_{ \hk_1+ \hk_2}
.
\end{align}
It follows then that the $N=2$ part of the EFT 
is
\begin{align}
S_{E\,[2]}
&=
\int \prod_{i=1}^2 \frac{d^D \hk_i}{(2\pi)^D}
\frac{1}{2!} \epsilon_{\mu_1}^{a}(\hk_1) 
\left( \delta^{\mu_1\mu_2} k_1^2  \delta_{\hk_1+\hk_2}
 \right) 
\epsilon_{\mu_2}^{a}(\hk_2)
. 
\end{align}
Comparing the previous expression with the EFT expressed using the 
canonical fields we get at this order in the number of fields \(A\)
\begin{equation}
v_{[2] i}=0,~~~~i=0,1,2
,~~~~
\xi=-\oh,~~~~
g_0=0,~~~~
r_{1}=r_{2}=0,
\end{equation}
and the gauge fixing action, always up to \(A^{2}\)
\begin{align}
\label{eq:gf_A^2}
S_{E\,[2], g.f.}
&=
\int d^D x
\left[ -\oh (\partial^\mu A_\mu^{a})^2 \right]
,
\end{align}
and no field redefinition is needed.
In order to describe how we proceed with interaction terms we have to
discuss what happens with Feynman vertices.

\subsection{Vertices and \Subvertices}
\label{sec:vert_and_subvert}
When we start with a field theory we can compute the Feynman vertices
and then compute Green functions 
by summing all the corresponding Feynman diagrams. 
Using these Green functions we can then compute the \(S\) matrix elements by
using the LSZ reduction formula which amounts to put on shell the
external legs after having truncated the legs.

In general given the part of the EFT action with \(N\) fields
\(\cL_{[N]}\) the
corresponding Feynman vertex can have up to \(N!\) terms since it is
built to be totally symmetric with respect the permutations of
equal fields.
For example in the case of the simplest \(\phi^{N}(x)\) colorless scalar theory
there is actually only \(1\) term in the vertex,
while in the case of Yang-Mills for $N=3$
we have $3!=6$ terms but for $N=4$ we have only $\oh 4!=12$ terms.

Consider a generic field \(\Phi_{A}(x)\) with $M$ components \(A=1,\dots
M\)  where \(A\) stands for both color and  space time indices.
Its polarization is then \(\Phi_{A}(k)\). 
The totally symmetric Euclidean vertex 
$V_{[N ]}\equiv V_{A_{1}\dots  A_{N}}(k_{1}\dots , k_{N})$
may have \(N!\) terms and it is defined by
\begin{equation}
-S_{E\,[N]}
= \int \prod_{i=1}^N \frac{d^D k_i}{(2\pi)^D}
\frac{1}{N!} 
V_{A_{1}\dots A_{N}}(k_{1}\dots , k_{N})
~\Phi_{A_{1}}(k_{1}) \dots \Phi_{A_{N}}(k_{N})
,
\end{equation}
where the momentum conservation 
\( (2\pi)^{D}\delta^{D}(\sum_{i} k_{i}) \equiv  \delta_{\sum_{i} k_{i}}\) 
is included into the definition of the vertex.

Because of the way we build the vertices 
a \(S\) matrix element with \(N\) fields may have \(N!\) terms
only from the vertex $V_{[N]}$. To these terms we must then add all
the others coming from connecting vertices with fewer legs.

\begin{figure}[hbt]
\begin{center}
\def\svgwidth{250px}
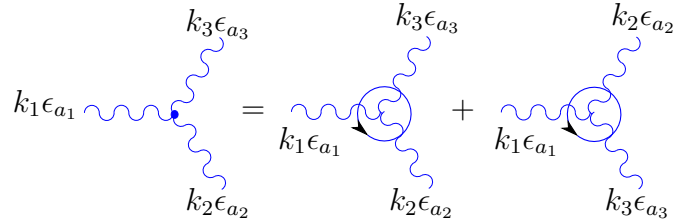
\end{center}
\vskip -0.5cm
\caption{
The 3 point totally symmetric vertex $V_{[3]}$ is given as a sum of two 
cyclically symmetric ones $V^{(123)}_{[3]}$ and $V^{(132)}_{[3]}$. 
}
\label{fig:V3}
\end{figure}

Nevertheless the comparison between open string theory and its EFT 
can be made easier if we
split the Feynman vertices into cyclically invariant \subvertices.
This split is shown in figure \ref{fig:V3} where the 3 gluon Feynman vertex
is written as the sum of two cyclically invariant
\subvertices{}  which are pictured with a circle with a direction.
Then we can compare one (out of \((N-1)!\))
string diagram with the corresponding color ordered Feynman diagram built
using the color ordered \subvertices.
In the case of the previous example with \(N=3\) this means comparing
the first string diagram on the rhs in figure \ref{fig:V3_string} with
the first Feynman sub-diagram on the rhs in figure \ref{fig:V3} (or
that is the same the second ones in the same figures).

\begin{figure}[hbt]
\begin{center}
\def\svgwidth{250px}
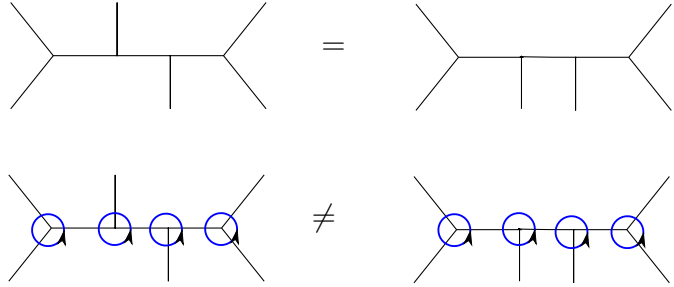
\end{center}
\vskip -0.5cm
\caption{
The two Feynman diagrams of the first line are equal because vertices
are totally symmetric  under permutations 
while the 2 (out of $2^4=16$) cyclically symmetric (color ordered) Feynman
diagrams in the second line differ.
}
\label{fig:F_vs_O_F}
\end{figure}

The same result applies when we compare Feynman diagrams involving
more than one vertices.
In general to a Feynman diagram build with $N_{3}$ 3 vertices corresponds
$2^{N_3}$ ordered Feynman diagrams.
For example in figure \ref{fig:F_vs_O_F} we show how a Feynman graph
built using the usual $3$ vertex can be drawn in many different ways
because of the permutation symmetry of the vertex.
Nevertheless using the cyclically symmetric vertex there is only one
way of drawing a graph. 
When we write all vertices in a Feynman diagram as sum of cyclically
symmetric  \subvertices{} and we expand this ``product'' we get a
\(1-1\) correspondence between 
these color ordered Feynman diagrams built using the ordered \subvertices{}
and the string color ordered diagrams.

Figure \ref{fig:V4_string-V3V3} shows what happens when we compare the
string diagrams of \(N=4\) gluons which have a pole in the \(s\) channel with the
corresponding Feynman diagram with a pole in the \(s\) channel.

\subsection{Dealing with interaction terms}
\label{sec:dealing_w_interactions}
Since we can compare color ordered string amplitudes with 
EFT color ordered Feynman diagrams built using the cyclically
symmetric \subvertices,
it is natural to try to read the cyclically invariant Feynman vertices
directly from string amplitudes with off shell/non physical states
which can be obtained from factorization.
This can be described in a more precise way.
In the case of $N=3$ we can read directly the $V^{(123)}_{[3]}$ while
for $N=4$ and greater $N$ we need first to subtract the poles and then
read the contact interactions.

However already for the \(N=3\) gluons case this does not work exactly.
It turns out to be possible to identify the $N=3$ gluon string amplitude
with the cyclically symmetric \subvertex{} up to gauge conditions,
i.e. up to terms proportional to \(\epsilon\cdot k\) as shown in
eq. (\ref{V3-gluons-symmetric+diff}).

This difference between the off shell string vertex and the EFT
cyclically invariant \subvertex{} is then at the origin 
of some contact terms in the quartic (and higher) coupling because of
the Ward identity. 
Moreover this difference causes a more annoying fact that 
it is not possible to compare off shell color ordered amplitudes but
only on shell ones, i.e. pieces of an \(S\) matrix element
\footnote{All these problems may perhaps be avoided using the twisted
  propagator which allows for cyclically invariant vertices}.
We will discuss this point in section \ref{sec:fact_N1}.

We read therefore the Feynman \subvertices{} as suggested by string
theory by mimicking it as close as possible with a \subvertex.
Then we can compute the totally symmetric Feynman vertices and 
compare these with the most general gauge fixed action.
It turns out that they cannot be derived directly from a gauge fixed
EFT written in terms of the canonical fields.
In fact the resulting vertices are written using fields which are not
the ones used to write the EFT but they are connected by
to them by a field redefinition.
Obviously one can use the canonical fields in the EFT but 
then the EFT vertices differ by more terms with respect to 
the string amplitudes.

\COMMENTO{
\subsection{A reminder on gauge fixing and gauge fermions}
\label{sec:gauge_fix_and_ferm}

Before proceeding in the stringy computations we would like to recall
very few facts on gauge fixing and gauge fermions in BRST formalism.

The point is that given a gauge fixing we can have many corresponding
gauge fermions as the existence of the \(\xi\) parameter shows.
Obviously different gauge fermions corresponds to different
propagators.

When we choose an off-shell extension of the 3 point amplitudes we
choose a piece of the gauge fixing.

Another piece is given by the propagator we choose. In fact string
chooses a well defined propagator but we can choose whichever
propagator we like.
Looking at factorization it seems that string chooses
\(P^{\mu\nu}(l)=\frac{\eta^{\mu\nu}}{\alpha' l^{2}} \)
while we could choose
\(P^{\mu\nu}(l)=\frac{\eta^{\mu\nu}- (1-\frac{1}{\xi})\frac{l^{\mu}
    l^{\nu}}{l^{2}}}
{\alpha' l^{2}} \).

The difference of gauge fixing shows up in f.x. 4 point effective
vertex which is the left over of the low energy limit of the 4 point
amplitude to which we subtract the Feynman diagrams corresponding to
the 3 point connected by the propagator.
} 

\section{String amplitudes: 3 points}
\label{sec:3points}
We would now implement in practice what we have discussed in the
previous section.
In particular we would like to determine the $3$ vertex suggested by
string theory and then find the gauge fixing and the field
redefinition necessary to map it to the EFT written the standard field.



\subsection{Three gluons amplitude}
\label{sec:3gluons}
It is standard matter (see for example \cite{GSW}) to compute the three photons
partial amplitude once we have given the photon vertex operator
\begin{equation}
V(x; \hk, \he)
= +\ii \he \cdot \partial \hX(x,x) e^{\ii \hk \cdot \hX(x,x)}
,
\end{equation}
where the hatted quantities are adimensional, for example
\(\hk=\sqtap k\) is the adimensional momentum.
We compute the partial amplitude 
not requiring that the in and out state be on shell or
transverse.
The reason is that this is what we see by factoring the \(5\) point amplitude.
The basic contribution to the truncated Euclidean Green function  is
then 
\begin{align}
A_{1^{*} 2 3^{*}}
&=
A(\hk_{1}^{*}, \he_{{1}}^{*}; \hk_{2},  \he_{{2}}; \hk_{3}^{*},
                   \he_{{3}}^{*})
=
\langle\langle \hk_1^{*}, \he_{1}^{*}|  
~V(x=1; \hk_2, \he_2)
~ |\hk_3^{*}, \he_{3}^{*}\rangle
\nonumber\\
&=
 \langle\langle \hk_1^{*}, 0| \he_1^{*} \cdot \alpha_1
~V(x=1; \hk_2, \he_2)
~ \he_3^{*} \cdot \alpha_{-1} |\hk_3^{*}, 0\rangle
\nonumber\\
&=
\langle\langle \hk_1^{*}, 0|  \he_1^{*} \cdot \alpha_1
~: \he_2 \cdot\left( \alpha_{-1}+\alpha_{0}+\alpha_{1}\right)
\ee^{i \hk_{2} x_{0} }
\ee^{\hk_2 \cdot \alpha_{-1} } 
\ee^{-\hk_2 \cdot \alpha_{1} } :
~\he_3^{*} \cdot \alpha_{-1}^{*} |\hk_3, 0\rangle
\nonumber\\
\label{V3-gluons-non-symmetric}
&=
[ 
- \he_{1}^{*}\cdot \he_{2} ~ \hk_{2}\cdot \he_{3}^{*}
+  \he_{3}^{*}\cdot\he_{1}^{*}~ \he_{2}\cdot \hk_{3}^{*} 
- \he_{1}^{*}\cdot \hk_{2} ~\he_{2}\cdot \hk_{3}^{*} ~\he_{3}^{*}\cdot \hk_{2}
+\he_{2}\cdot \he_{3}^{*} ~\hk_{2}\cdot\he_{1}^{*}
]
\delta_{\hk_{1}+\hk_{2}+\hk_{3}}
\\
%
\label{V3-gluons-symmetric+diff}
&=
[ - \he_{1}^{*}\cdot \he_{2} ~\hk_{2}\cdot\he_{3}^{*} 
- \he_{2}\cdot \he_{3}^{*} ~\hk_{3}^{*}\cdot\he_{1}^{*} 
- \he_{3}^{*}\cdot \he_{1}^{*} ~\hk_{1}^{*}\cdot\he_{2} 
\nonumber\\
&
\phantom{= [} 
+ \he_{1}^{*}\cdot \hk_{2} ~\he_{2}\cdot \hk_{3}^{*} ~\he_{3}^{*}\cdot  \hk_{1}^{*} 
\nonumber\\
&\phantom{= [}
+ \he_{1}^{*}\cdot \he_{2} ~\hk_{3}^{*}\cdot\he_{3}^{*}
+ \he_{1}^{*}\cdot \hk_{2} ~\he_{2}\cdot \hk_{3}^{*} ~\he_{3}^{*}\cdot  \hk_{3}^{*} 
]
\delta_{\hk_{1}+\hk_{2}+\hk_{3}}
,
\end{align}
where the \(^{*}\) means that the corresponding starred quantity may not satisfy the physical conditions.
It is the previous expression properly normalized, i.e.
\(\cCz \cNz^{3 }A_{1^{*} 2 3^* }\) that we want to
mimic with the \subvertex{} of the EFT.
Few things are worth noticing.
First eq. (\ref{V3-gluons-non-symmetric}) is antisymmetric in the
exchange of the two non physical gluons \(1\) and \(3\). This makes
impossible to interpret it as a piece of a usual EFT since the
Feynman vertices are totally symmetric in the exchange of gluons.
Secondly the last way of writing the partial amplitude 
\(A_{1^{*}   2 3^* }\) in eq. (\ref{V3-gluons-symmetric+diff}) 
shows that 
the amplitude is cyclically invariant when we use the
  gauge condition \(\he_{3}^{*}\cdot k_{3}= 0\) for the third state, i.e.
\(A_{1^{*} 2 3^* }\) is cyclically invariant when $\he_3^{*}$ is
transverse but eventually off shell, since then the last line vanishes.
Obviously
it is possible to write an analogous expression where we require the
transversality for the first state \(\he_{1}\cdot \hk_{1}= 0\).

Only when all states are physical, i.e. on shell and transverse 
the amplitude has  on shell gauge invariance,
i.e. it is invariant under \(\hepsilon\rightarrow \hepsilon + \hk \) with 
\(\hk^{2}=0\).

The  \(S\) matrix element from string theory for non abelian gluons
can then be obtained from the \amplitude{} as
\begin{eqnarray}
  \label{eq:full-3pt}
  \cA_{1 2 3}(\hk_{1}, \hepsilon_{\mu_{1}a_{1}}; 
\hk_{2}, \hepsilon_{\mu_{2}a_{2}}; \hk_{3}, \hepsilon_{\mu_{3}a_{3}}) 
&=& 
\cCz \cNz^3 
\left[
A_{1 2 3} ~tr(T_{a_1}T_{a_2}T_{a_3})
+A_{1 3 2} ~tr(T_{a_1}T_{a_3}T_{a_2})
\right]
\nonumber\\
\end{eqnarray}
and it is obtained by taking all states physical, 
substituting the abelian polarizations
\(\hepsilon_{i}\) with their non abelian ones \(\hepsilon_{a_{i}}\) and
multiplying by the Chan-Paton factors, explicitly in the previous
expression we have 
\begin{equation}
A_{1 2 3} = 
A(\hk_{1}, \hepsilon_{a_{1}}; \hk_{2},  \hepsilon_{a_{2}}; 
\hk_{3},  \hepsilon_{a_{3}})
,
\end{equation}
and there is no summation over the color indices.
The full amplitude is depicted in figure \ref{fig:V3_string}.

Now because of the on shell condition \(\hk_{i}^{2}=0\) it follows that
all the momenta \(\hk_{i}\) are parallel 
as can be easily seen since on shell \(\hk_{i} \cdot \hk_{j}=0\) and we
can choose any \(\hk\) in the light cone direction.
Therefore both the \amplitude{} and the \(S\) matrix vanish
\begin{eqnarray}
  \label{eq:full-3pt=0}
  S_{1 2 3}(\hk_{1}, \hepsilon_{\mu_{1}a_{1}}; 
\hk_{2}, \hepsilon_{\mu_{2}a_{2}}; \hk_{3}, \hepsilon_{\mu_{3}a_{3}}) 
&=&0
.
\end{eqnarray}

\subsection{The general three gluons up to three 
derivatives Lagrangian}
In order 
to reconstruct the gauge fixed EFT from the previous \(S\) matrix
we write down the most general Lagrangian with \(3\) gluons and up to
\(3\) derivatives.
From the Lagrangian we deduce the \(3\) Feynman vertex and then we require
that it yields a 3 point $S$ matrix element vanishing on shell.
Besides this constraints we have nevertheless to respect the pole
structure of the 4 and higher point \(S\) matrix amplitudes, i.e. given the 4 point
\(S\) matrix amplitude the result of subtracting the contribution from the
reducible Feynman diagrams obtained by joining two 3 point vertices
must be pole free\footnote{
This requirement is not true when dealing with Green functions as
we show in section \ref{sec:fact_N1} since the the stringy off shell amplitude
cannot be interpreted as a piece of a usual Feynman vertex.
}.

Nevertheless
as discussed in the previous section \ref{sec:dealing_w_interactions}
our main idea is to proceed in a different way and
we use the off shell extension \(\cCz \cNz^{3 }A_{1^{*} 2 3^* }\) 
to read the 3 vertex suggested by string theory for a EFT.
However we consider the most general Lagrangian in order to
discuss how the string choice minimizes the number of terms in the
\(3\) and \(4\) point vertices.

The general cubic  effective action with up to three derivatives reads\footnote{
The easiest way to obtain it is to work in momentum space.
The terms with one momentum are immediate to find.
The terms with three momenta fall into two categories either
$(\epsilon\cdot k)^3$ or $ (\epsilon\cdot\epsilon) (\epsilon\cdot k) 
(k \cdot k)$.

Let us consider the first class.  Using cyclicity we have $3^3$ terms
$\epsilon_1\cdot k_i~\epsilon_2\cdot k_j~\epsilon_3\cdot k_l $ since $i,j,k\in\{1,2,3\}$.
Using momentum conservation we can consider only $2^3$ terms,
i.e. those with $i\ne 1, j\ne 2, l\ne 3$.
Then using again cyclicity we are left with $4$ terms, those with
coefficients $c_3,\dots c_6$ in eq. (\ref{eq:3pts_offshell_generic_part}).
An example of the use of cyclicity is the fact that 
the term with $(i,j,k)=(2,3,2)$ is equivalent to $(i,j,k)=(2,1,1)$. 

Now consider the second class.
Using cyclicity we have $3^3$ terms like
$\epsilon_1\cdot\epsilon_2~\epsilon_3\cdot k_l~ k_i\cdot k_j $.
Again momentum conservation allows us to consider the cases $i,j,l\ne 3$.
They are $6$ and are the terms with coefficients $c_7,\dots c_{12}$ in eq. (\ref{eq:3pts_offshell_generic_part}).
}
\begin{align}
S_{E\,[3]} = \int d^D x\,
\frac{1}{{\kappa}}
tr
[
&+c_1  \partial_\mu A_\nu\, A_\mu\, A_\nu
&
&+c_2  \partial_\mu A_\nu\, A_\nu\, A_\mu
\nonumber\\
&+c_3  \partial_\mu A_\nu\, \partial_\nu A_\lambda\, \partial_\lambda A_\mu
&
&+c_4  \partial_\lambda \partial_\nu A_\mu\, \partial_\mu A_\nu\, A_\lambda
\nonumber\\
&+c_5  \partial_\lambda \partial_\nu A_\mu\, \partial_\mu A_\nu\, A_\lambda
&
&+c_6  \partial_\nu A_\mu\, \partial_\lambda \partial_\mu A_\nu\, A_\lambda 
\nonumber\\
&+c_7  \partial^2 \partial_\lambda A_\mu\, A_\mu\, A_\lambda
&
&
+c_8  \partial_\lambda A_\mu\, \partial^2 A_\nu\, A_\lambda
\nonumber\\
&+c_9 \partial_\rho \partial_\lambda A_\mu\,\partial_\rho A_\nu\, A_\lambda
&
&
\nonumber\\
&+c_{10} \partial^2 A_\mu\, \partial_\lambda A_\mu \, A_\lambda
&
&
+c_{11}  A_\mu\, \partial^2\partial_\lambda A_\mu \, A_\lambda
\nonumber\\
&+c_{12} \partial_\rho A_\mu\,\partial_\rho \partial_\lambda A_\nu\, A_\lambda
&
&
]
.
\end{align}
Notice that all these terms give a vanishing 3 point \(S\)
matrix. This can be more easily looking at the corresponding Feynman
vertex in eq.s (\ref{eq:3pts_offshell_generic},
\ref{eq:3pts_offshell_generic_part}). 
In particular it is necessary to remember that all \(k_{i}\) are
parallel on shell and hence \(\epsilon_{i} \cdot k_{j}=0\).

Interpreting this cubic interaction as coming from a gauge invariant
action with with a non linear gauge fixing as eq. (\ref{eq:S_g.f.})
and a field redefinition as in eq. (\ref{eq:field_red}) (assuming a
canonical kinetic term which implies \(g_{0} =v_{[2]1} =v_{[2]2} =0\))
requires\footnote{
The dependence of coefficients \(c_{1}\dots c_{4}\) on \(g\) and
\(v_{[3] 0}\) can be immediately read by
expanding the Lagrangian, the other requires a little more work.
}
\COMMENTOBUH{ Changed signs for \(g\) terms in \(c_{1,2}\) passing to Euclidean}
\begin{align}
\label{eq:cs_in_L[3]}
c_1&=-i g - 2 g_1,~~~~
c_2=+i g - 2 g_1, 
\nonumber\\
c_3&= +v_{[3] 0} - 2 g_2,~~~~
c_4= -v_{[3] 0} - 2 g_2,
\nonumber\\
c_5&= -2 g_3 - 6  g_2,~~~~
c_6= -2 g_3 - 6  g_2,
\nonumber\\
c_7&=-2 g_5 - r_3 - r_5,~~~~
c_{10}= -3 v_{[3] 0}  -2 g_5 - r_3 - r_4 + r_5, 
\nonumber\\
c_8&= 3 v_{[3] 0}  -2 g_6 + r_3 + r_4 - r_5, ~~~~
c_{11}= - 2 g_6 + r_3 + r_5,
\nonumber\\
c_9&=  3 v_{[3] 0} - 2  g_4 - 2 r_5,~~~~
c_{12} = - 3 v_{[3] 0} - 2 g_4 + 2 r_5 
.
\end{align}
In particular the previous vertex can {\bf not} become 
the usual three vertex in the linear Lorentz gauge unless
\(c_{2}= - c_{1}\),  \(3 c_{3} = - 3 c_{4} = c_{8} = c_{9} = -c_{10} =
- c_{12}\) and $c_{5, 6, 7, 11}=0$.   
This happens because the usual three vertex involves 
the commutator of the algebra elements 
\(tr( T_{a} [ T_{b}, T_{c}]) \) which is totally antisymmetric in the
exchange of \(a, b, c\).
When these conditions are not satisfied the cubic interaction does not
originate from a gauge invariant action with linear gauge fixing
and we must interpret it as originating from a gauge fixed action with
non linear gauge fixing and a field redefinition.

The previous cubic interaction gives raise to the Euclidean Feynman cubic
vertex defined by
\begin{equation}
-S_{E\,[3]}
= \int \prod_{i=1}^3 \frac{d^D k_i}{(2\pi)^D}
\frac{1}{3!} 
V_{\mu_{1} a_{1}, \mu_{2} a_{2}, \mu_{3} a_{3} }(k_{1},  k_{2},  k_{3})
~\epsilon^{\mu_{1}}_{a_{1}}(k_{1}) \epsilon^{\mu_{2}}_{a_{2}}(k_{2}) 
\epsilon^{\mu_{3}}_{a_{3}}(k_{3})
.
\end{equation}
As discussed in section \ref{sec:vert_and_subvert} it is convenient to write
this cubic vertex as the sum of two cyclically
invariant  \subvertices{}  as shown in figure \ref{fig:V3} as
\begin{eqnarray}
  \label{eq:3pts_offshell_generic}
V_{\mu_{1} a_{1}, \mu_{2} a_{2}, \mu_{3} a_{3} }(k_{1},  k_{2},  k_{3})
=&
\frac{1 }{\kappa}
\Bigl[
V^{(1 2 3)}_{\mu_{1};~\mu_{2};~\mu_{3}}(k_{1}, k_{2}, k_{3})  
~tr(T_{a_1}T_{a_2}T_{a_3}) &
\nonumber\\
&+
V^{(1 2 3)}_{\mu_{1};~\mu_{3};~\mu_{2}}(k_{1}, k_{3}, k_{2})  
~tr(T_{a_1}T_{a_3}T_{a_2}) &
\Bigr]
,
\end{eqnarray}
where\footnote{
The coefficients $3 c_3$  and $3 c_4$ come from the fact that the corresponding
structures are cyclically symmetric.
The different signs from the different momentum powers $i k$ vs $(i k)^3$.
}
\COMMENTOBUH{ Changed overall sign passing to Euclidean}
\begin{eqnarray}
  \label{eq:3pts_offshell_generic_part}  
  V^{(1 2 3)}_{\mu_{1};~ \mu_{2};~\mu_{3}}
&
\epsilon^{\mu_{1}}_{a_{1}}\epsilon^{\mu_{2}}_{a_{2}}\epsilon^{\mu_{3}}_{a_{3}}
=
(+\ii)
\Bigl[
- c_{1} ( 
 \epsilon_{a_{1}}\cdot \epsilon_{a_{2}} ~\epsilon_{a_{3}} \cdot k_{2} 
+\mbox{cycl}
)
-c_{2} (
 \epsilon_{a_{1}}\cdot \epsilon_{a_{2}} ~\epsilon_{a_{3}}\cdot k_{1} 
+\mbox{cycl}
)
\nonumber\\
&
+ 3 c_{3} ~\epsilon_{a_{1}}\cdot k_{2}  ~\epsilon_{a_{2}}\cdot k_{3}
  ~\epsilon_{a_{3}}\cdot k_{1}
+ 3 c_{4} ~\epsilon_{a_{1}}\cdot k_{3} ~\epsilon_{a_{2}}\cdot k_{1}
  ~\epsilon_{a_{2}}\cdot k_{2} 
\nonumber\\
&
+ c_{5} (\epsilon_{a_{1}}\cdot k_{2}  ~\epsilon_{a_{2}}\cdot k_{1}
  ~\epsilon_{a_{3}}\cdot k_{1}
+\mbox{cycl}
)
+ c_{6} (\epsilon_{a_{1}}\cdot k_{2} ~\epsilon_{a_{2}}\cdot k_{1}
  ~\epsilon_{a_{2}}\cdot k_{2} 
+\mbox{cycl}
)
\nonumber\\
&
+ c_{7} 
(\epsilon_{a_{1}}\cdot\epsilon_{a_{2}}~ \epsilon_{a_3}\cdot k_{1}~k_{1}^2
+\mbox{cycl}
)
+ c_{8} 
(\epsilon_{a_{1}}\cdot\epsilon_{a_{2}}~ \epsilon_{a_3}\cdot k_{1}~k_{2}^2
+\mbox{cycl}
)
\nonumber\\
&
+ c_{9} 
(\epsilon_{a_{1}}\cdot\epsilon_{a_{2}}~ \epsilon_{a_3}\cdot
  k_{1}~k_{1}\cdot k_2
+\mbox{cycl}
)
\nonumber\\
&
+ c_{10} 
(\epsilon_{a_{1}}\cdot\epsilon_{a_{2}}~ \epsilon_{a_3}\cdot k_{2}~k_{1}^2
+\mbox{cycl}
)
+ c_{11} 
(\epsilon_{a_{1}}\cdot\epsilon_{a_{2}}~ \epsilon_{a_3}\cdot k_{2}~k_{2}^2
+\mbox{cycl}
)
\nonumber\\
&
+ c_{12} 
(\epsilon_{a_{1}}\cdot\epsilon_{a_{2}}~ \epsilon_{a_3}\cdot
  k_{2}~k_{1}\cdot k_2
+\mbox{cycl}
)
\Big]
\delta_{k_{1}+k_{2}+k_{3}}
.
\nonumber\\
\end{eqnarray}
Matching the structure of cyclical \subvertex{} as close as possible
to the off shell amplitude (\ref{V3-gluons-symmetric+diff}) gives
\COMMENTOBUH{ Changed $g$ sign passing to Euclidean}
\begin{align}
c_1=  -2 \ii g,&~~~~
c_2=0,
\nonumber\\
c_3= 2 v_{[3] 0},&~~~~
c_4=0,
\nonumber\\
c_{5, 6, 7, 8, 9, 10, 11, 12}&=0
\nonumber\\
g_{1}&= \oh \ii g,
\nonumber\\
g_{2}= - \oh v_{[3] 0},&~~~~
g_{3}=-3 g_{2}
\nonumber\\
g_{4}= \frac{3}{2} v_{[3] 0},&~~~~
g_{5}=-g_{6}=r_{3}+r_{5}
\nonumber\\
r_{4}=0,&~~~~
r_{5} = \frac{3}{2} v_{[3] 0}
.
\end{align}
A rapid look to eq.s (\ref{eq:cs_in_L[3]}) reveals that these
coefficients cannot be reproduced simply using a gauge fixing and that
we therefore need a field redefinition.
We find the gauge fixed Lagrangian
\COMMENTOBUH{ Changed $g$ sign passing to Euclidean}
\begin{align}
S_{E\,[3] gauge\,fixed}
=
\int d^D x\,
\frac{1}{{\kappa}}
tr \Bigl(
-2 i g  \partial_\mu A_\nu\, A_\mu\, A_\nu
+2 v_{[3] 0}  \partial_\mu A_\nu\, \partial_\nu  A_\lambda\, \partial_\lambda A_\mu
\Bigr)
,
\end{align}
the gauge fixing Lagrangian
\COMMENTOBUH{ Changed $g$ sign passing to Euclidean}
\begin{align}
S_{E\,gf} = \int d^D x\,
&
\frac{\xi}{{\kappa}}
tr 
\Bigl(
\partial_\mu A_\mu
+\ii \frac{ g}{2 \xi} A_\mu A_\mu
\nonumber\\
&
-\frac{ v_{[3] 0} }{2 \xi} \partial_\mu A_\mu \partial_\nu A_\nu
+ \frac{ 3 v_{[3] 0} }{2 \xi} \partial_\nu A_\mu \partial_\mu A_\nu
- \frac{ v_{[3] 0}  + 2 r_{3} }{2 \xi} [\partial^2 A_\mu,  A_\mu]
+\dots \Bigr)^2
,
\end{align}
with \(\xi=-\oh\) as from eq. (\ref{eq:gf_A^2})
and the field redefinition
\begin{align}
A_\mu=
& \rA_\mu 
+r_3 [\rA_\mu,  \partial_\rho\rA_\rho]
+ \frac{ 3 v_{[3] 0} }{2 \xi} [\rA_\rho,  \partial_\mu\rA_\rho]
+\dots
.
\end{align}
If we want to match also the coefficient we need to match the previous
\subvertex{} with \(\cCz \cNz^{3 }A_{1^{*} 2 3^* }\)  and set
\COMMENTOBUH{ Changed $g$ sign passing to Euclidean}
\begin{align}
\label{eq:c1,3 and C0}
c_{1}=-2 \ii g = -\ii \cCz \cNz^{3} \tap^{2-\oh D},~~~~
c_3= 2 v_{[3] 0}= -\frac{1}{3} \ii \cCz \cNz^{3} \tap^{3 -\oh D}
,
\end{align}
thus finding the usual result
\begin{equation}
v_{[3] 0}= - \frac{1}{3} \ii \tap  g 
.
\end{equation}

If we do not want to use field redefinition we have more possibilities
on the closest possible vertex has gauge fixed Lagrangian.
One possibility is given by the gauge fixed Lagrangian
\COMMENTOBUH{ Changed $g$ sign passing to Euclidean}
\begin{align}
S_{E\,[3] gauge\,fixed}
=
\int d^D x\,
\frac{1}{{\kappa}}
tr \Bigl(
-2 i g  \partial_\mu A_\nu\, A_\mu\, A_\nu
+2 v_{[3] 0}  \partial_\mu A_\nu\, \partial_\nu  A_\lambda\, \partial_\lambda A_\mu
+ 3 v_{[3] 0} \partial_\mu A_\nu\, [\partial_\lambda A_\mu,\, \partial_\lambda A_\nu]
\Bigr)
,\end{align}
and the gauge fixing Lagrangian
\COMMENTOBUH{ Changed $g$ sign passing to Euclidean}
\begin{align}
S_{E\,gf} = \int d^D x\,
&
\frac{\xi}{{\kappa}}
tr 
\Bigl(
\partial_\mu A_\mu
+\ii \frac{ g}{2 \xi} A_\mu A_\mu
-\frac{ v_{[3] 0} }{2 \xi} \partial_\mu A_\mu \partial_\nu A_\nu
+ \frac{ 3 v_{[3] 0} }{2 \xi} \partial_\nu A_\mu \partial_\mu A_\nu
+\dots \Bigr)^2
.
\end{align}
Another possibility is given by the gauge fixed Lagrangian
\COMMENTOBUH{ Changed $g$ sign passing to Euclidean}
\begin{align}
S_{E\,[3] gauge\,fixed}
=
\int d^D x\,
\frac{1}{{\kappa}}
tr \Bigl(
-2 i g  \partial_\mu A_\nu\, A_\mu\, A_\nu
+v_{[3] 0}  F_{\mu \nu} F_{\nu \lambda} F_{\lambda \mu}
\Bigr)
,\end{align}
and the gauge fixing Lagrangian
\COMMENTOBUH{ Changed $g$ sign passing to Euclidean}
\begin{align}
S_{E\,gf} = \int d^D x\,
&
\frac{\xi}{{\kappa}}
tr 
\Bigl(
\partial_\mu A_\mu
+\ii \frac{ g}{2 \xi} A_\mu A_\mu
+\dots \Bigr)^2
.
\end{align}

\subsection{The abelian limit and and intuitive explanation of the
  Gervais-Neveu gauge}

Looking to the possible terms in the \subvertex{} \(V^{(123)}\) it is clear that
some om them become exchanged under non cyclical permutations.
For example \(c_{1}\) and \(c_{2}\) are exchanged when 
\(1 \leftrightarrow 2\).
In more formal way \( c_{2}^{(123)}=c_{1}^{(213)}\).
This means that we can know \(c_{2}^{(123)}\) if we know
\(c_{1}^{(123)}\) since by exchanging \(1 \leftrightarrow 2\) we can
compute \(c_{1}^{(213)}\).
Therefore \(c_{2}^{(123)}\) is redundant and can be likely set to zero by
choosing a gauge.
In facts string theory chooses \(c_{2}=0\) (or equivalently \(c_{1}=0\)).
It seems that 
string theory be choosing the minimal number of terms from which we
can reconstruct both the abelian and non abelian theory.
Because of this also the abelian theory has non vanishing \(3\) vertex.

\section{Four gluons amplitude, propagator and contact terms}
\label{sec:4points}
The basic partial amplitude (and not correlator since this is already the
integrated correlator) is%
\footnote{Notice that this expression is
naive since it is divergent as it stands because of the sum over
infinite intermediate states (this divergence seemed to be well known in
1971, see \cite{Alessandrini:1971fx} after eq. 4.40).   
This is easily seen in the four tachyons amplitude 
\( \int_{0}^{1} dx~x^{\hk_{3} \cdot \hk_{4} }
(1-x)^{\hk_{2}\cdot\hk_{3}}\)
where the term \((1-x)^{\hk_{2}\cdot\hk_{3}}\) can be expanded around
\(x=0\) inside the integral and this gives the \(s\) channel poles
Nevertheless the infinite summation cannot be exchanged with
the integral because the series is not uniformly convergent.

To give a proper meaning we need to use a regularized propagator as 
\(\Delta_r(\epsilon)= e^{-\epsilon  N}/(L^{(X)}_{0}-1) \) 
as well as consider a contribution from the $A_{2 3 4 1}$ amplitude
like what happens in string field theory where the infinite sum is
naturally performed.
For the time being we do not consider this and take the previous
expression as the integral of a correlator which is well defined.
}
\begin{align}
A_{1 2 3 4}
&=
A(\hk_1, \hepsilon_1;\dots \hk_4, \hepsilon_4) =
\langle\langle \hk_1, \hepsilon_{1}|  
~V(x=1; \hk_2, \hepsilon_2)
~\frac{1}{ L^{(X)}_{0}-1}
~V(x=1; \hk_3, \hepsilon_4)
~ |k_4, \hepsilon_{4}\rangle
.
\end{align}
\begin{figure}[hbt]
\begin{center}
\def\svgwidth{350px}
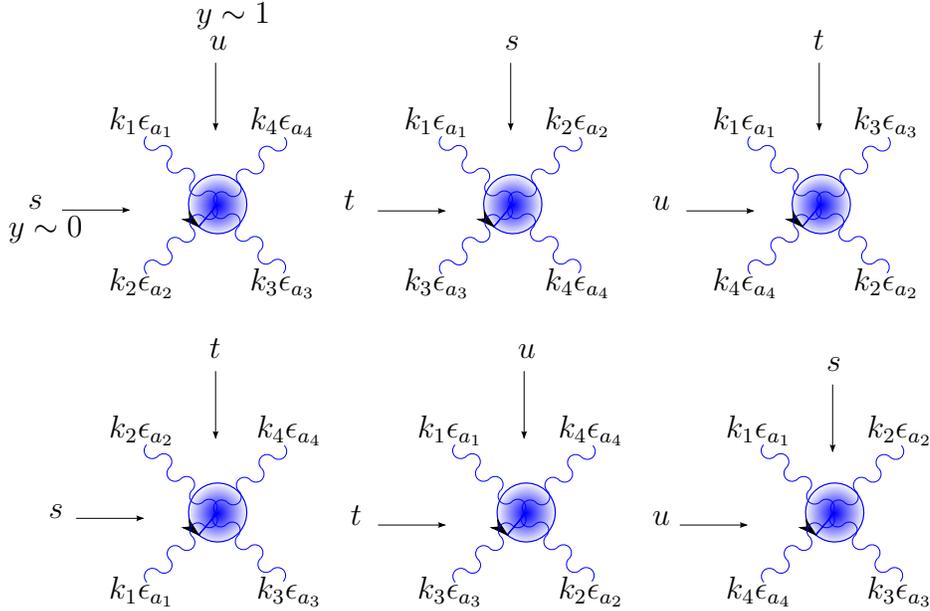
\end{center}
\vskip -0.5cm
\caption{
The six diagrams contributing to the $N=4$ amplitude with the
indication of the channels to which each diagram contributes.
The horizontal channel is the obvious one from the old way of writing
the amplitude. 
The vertical channel is the obvious one when using the
cyclicity of the amplitude.
}
\label{fig:V4_string}
\end{figure}
The full \(S\) matrix is then obtained from (see figure \ref{fig:V4_string})
\begin{align}
\cA_{1 2 3 4} 
= 
\frac{\alpha'}{\kappa}
\cCz \cNz^4 \Big\{ 
&
[ A_{1 2 3 4} ~tr(T_{a_1}T_{a_2}T_{a_3}T_{a_4}) + A_{1 2 4 3}~tr(T_{a_1}T_{a_2}T_{a_4}T_{a_3}) ]
\nonumber\\
+&
[ A_{1 3 4 2} ~tr(T_{a_1}T_{a_3}T_{a_4}T_{a_2}) + A_{1 3 2 4} ~tr(T_{a_1}T_{a_3}T_{a_2}T_{a_4})]
\nonumber\\
+&
[ A_{1 4 2 3 } ~tr(T_{a_1}T_{a_4}T_{a_2}T_{a_3}) + A_{1 4 3 2}
~tr(T_{a_1}T_{a_4}T_{a_3}T_{a_2})]
\Big\}
,
\end{align}
where we  substitute the abelian polarizations
\(\hepsilon_{i}\) with their non abelian ones \(\hepsilon_{i a_{i}}\).
In the previous equation the first line gives poles 
in the $s$ and \(u\) channels, 
the second to the \(s\) and $t$ ones 
and the last to the \(t\) and $u$ ones  where we defined
\begin{equation}
s=-(k_1+k_2)^2,~~~~
t=-(k_1+k_3)^2,~~~~
u=-(k_1+k_4)^2
.
\end{equation}

\subsection{Factorizing the $N=4$ amplitude on the gluons  and
  constraints on the \(c_{i}\) coefficients}
\label{sec:fact_N1}
In order to discuss how the string minimize the number of terms in the
vertices 
we would now find the constraints on the constants
\(c_{1,\dots 12}\) which arise in order to cancel the
physical poles.
In the following subsection we use these constraints to show that the
string solution is minimal in ensuing that the \(4\) vertex has the
minimal number of terms.

\begin{figure}[hbt]
\begin{center}
\def\svgwidth{350px}
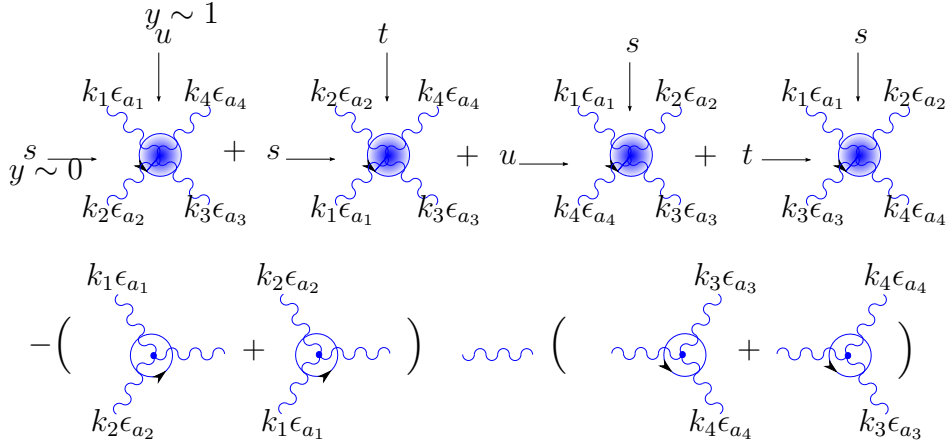
\end{center}
\vskip -0.5cm
\caption{
The ordered string diagrams with poles in the $s$ channel and the
Feynman diagram with a pole in the same channel.
To any ordered  string diagram corresponds a part of the Feynman
diagram computed with the ordered Feynman vertices.
}
\label{fig:V4_string-V3V3}
\end{figure}

\begin{figure}[hbt]
\begin{center}
\def\svgwidth{250px}
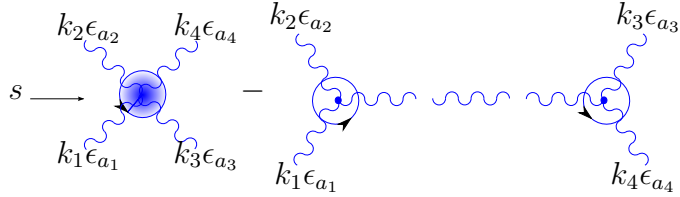
\end{center}
\vskip -0.5cm
\caption{
Single diagram subtraction
}
\label{fig:V4_string-V3V3_1diag}
\end{figure}

The cancellation of poles  can be checked 
by comparing the ordered string diagrams with a pole
in the $s$ channel (all the other channels would do the same)  
with the Feynman diagram from EFT which has a pole in
the same $s$ channel.
In order to do so we must see which of the six terms has a pole in the
$s$ channel.
It is obvious that $A_{1 2 3 4}$ and  $A_{1 2 4 3}$ have such a pole
but because of the cyclicity also $A_{1 3 4 2 }\equiv A_{2 1 3 4}$
and $A_{1 4 3 2 }\equiv A_{2 1 4 3}$ have therefore 
\begin{align}
\label{s_ch_string_N4_gluons}
\cA_{1 2 3 4} 
\sim_{s\rightarrow 0} 
\frac{\alpha'}{\kappa}
\cCz \cNz^4 \Big\{ 
&
[ A_{1 2 3 4} ~tr(T_{a_1}T_{a_2}T_{a_3}T_{a_4}) 
+ A_{1 2 4 3}~tr(T_{a_1}T_{a_2}T_{a_4}T_{a_3}) ]
\nonumber\\
+&
[ A_{1 3 4 2} ~tr(T_{a_1}T_{a_3}T_{a_4}T_{a_2}) ] +
[A_{1 4 3 2} ~tr(T_{a_1}T_{a_4}T_{a_3}T_{a_2})] + O(1)
\Big\}
.
\end{align}
To these ordered diagrams corresponds the Feynman diagram
\begin{align}
\label{s_ch_N4_gluons}
\epsilon^{\mu_{1}}_{a_{1}}\epsilon^{\mu_{2}}_{a_{2}}
~\frac{1}{\kappa}
&\Bigl[
V^{(1 2 3)}
_{\mu_{1};~\mu_{2};~\mu}(k_{1}, k_{2}, \qs)  
~tr(T_{a_1} T_{a_2} T_{b}) 
+
V^{(1 3 2)}
_{\mu_{1};~\mu;~\mu_{2}}(k_{1}, \qs, k_{2})  
~tr(T_{a_1} T_{b} T_{a_2}) 
\Bigr]
\nonumber\\
\times
&\frac{ \delta^{b c}  P(\qs)^{\mu\nu} }{\qs^2}
\nonumber\\
\times
\frac{1}{\kappa}
&\Bigl[
V^{(1 2 3)}
_{\nu;~\mu_{3};~\mu_{4}}(-\qs, k_{3}, k_{4})  
~tr(T_{c} T_{a_3} T_{a_4}) 
+
V^{(1 3 2)}
_{\nu;~\mu_{4};~\mu_{3}}(-\qs, k_{4}, k_{3})  
~tr(T_{c} T_{a_4} T_{a_3}) 
\Bigr]
~\epsilon^{\mu_{4}}_{a_{4}}\epsilon^{\mu_{3}}_{a_{3}}
\nonumber\\
\times
&
\delta_{\sum k_{i}}
,
\end{align}
with $k_1+k_2+\qs=-\qs+k_3+k_4=0$.
The  request is then that the expression
(\ref{s_ch_string_N4_gluons}) and (\ref{s_ch_N4_gluons}) have the same pole.
As shown in figure \ref{fig:V4_string-V3V3} and discussed above in
section \ref{sec:dealing_w_interactions}
the computation can be
simplified since to any ordered string diagram corresponds a piece of
the Feynman diagram built using the cyclically symmetric \subvertices.
Because of this we only need to compute the expression graphically
depicted in figure \ref{fig:V4_string-V3V3_1diag}.
Then the expression which corresponds to figure this
is given by
\begin{align}
\label{s_ch_subtraction}
\frac{\alpha'}{\kappa}&\cCz \cNz^4 A_{1 2 3 4} ~tr(T_{a_1}T_{a_2}T_{a_3}T_{a_4})
\nonumber\\
&-
\frac{1}{\kappa}
\epsilon^{\mu_{1}}_{a_{1}}\epsilon^{\mu_{2}}_{a_{2}}
V^{(1 2 3)}
_{\mu_{1};~\mu_{2};~\mu}(k_{1}, k_{2}, \qs)  
~tr(T_{a_1} T_{a_2} T_{b}) 
\frac{ \delta^{b c} \delta^{\mu\nu} }{\qs^2}
\frac{1}{\kappa}
V^{(1 2 3)}
_{\nu;~\mu_{3};~\mu_{4}}(-\qs, k_{3}, k_{4})  
~\epsilon^{\mu_{3}}_{a_{3}}\epsilon^{\mu_{4}}_{a_{4}}
~tr(T_{c} T_{a_3} T_{a_4})
\nonumber\\
&
\phantom{- ~ }
\times
\delta_{\sum k_{i}}
\\
\label{eq:s_ch_subtr_semplified}
=
\alpha'&\cCz \cNz^4 A_{1 2 3 4} 
-
\Bigl[
\epsilon^{\mu_{1}}_{a_{1}}\epsilon^{\mu_{2}}_{a_{2}}
V^{(1 2 3)}
_{\mu_{1};~\mu_{2};~\mu}(k_{1}, k_{2}, \qs)  
\frac{\delta^{\mu\nu} }{\qs^2}
V^{(1 2 3)}
_{\nu;~\mu_{3};~\mu_{4}}(-\qs, k_{3}, k_{4})  
~\epsilon^{\mu_{3}}_{a_{3}}\epsilon^{\mu_{4}}_{a_{4}}
\Bigr]
\delta_{\sum k_{i}}
\nonumber\\
&
\phantom{- ~ }
\frac{1}{\kappa} tr(T_{a_1}T_{a_2}T_{a_3}T_{a_4})
,
\end{align}
where we have already used the first suggestion which comes from
string, i.e. to use the propagator in Feynman gauge\footnote{
This does not mean that the gauge fixing is the usual Lorentz gauge
but only that the linear part of the gauge fixing is the usual Lorentz
gauge.}.
We have also used
\begin{align}
tr(X\, T_a)\,\delta^{a b}\,tr(T_b\,Y)
=\kappa\,tr(X\,Y)
~~~~
X,Y\in u(N)
.
\end{align}

The pole in the \(s\) channel of the string partial amplitude can be exposed by
simply inserting twice the unity at level \(N=1\) in the string
amplitude $A_{1 2 3 4}$ and get
\begin{align}
\label{eq:s_pole_A1234}
A_{1 2 3 4}&\sim_{s\rightarrow 0}
\int_{\qs}
\langle\langle \hk_1, \hepsilon_{1}|  
~V(x=1; \hk_2, \hepsilon_2)
\alpha^{\mu}_{-1}|\qs\rangle
~\frac{\delta_{\mu\nu}}{\alpha' \qs^{2} }
\langle\langle \qs| \alpha^{\nu}_{1}
~V(x=1; \hk_3, \hepsilon_4)
~ |\hk_4, \hepsilon_{4}\rangle
.
\end{align}
Comparing this expression with the EFT one in
eq. (\ref{s_ch_subtraction}) suggests to set 
\footnote{At first sight the choice of $\sqrt{-\cCz \cNz^4 }$ seems
quite odd and the choice $\sqrt{+\cCz \cNz^4 }$ would seem more natural
but it is the proper one when considering the results of the
comparison of the 3 vertex \ref{eq:c1,3 and C0}.
}
\begin{align}
\sqrt{-\cCz \cNz^4 }
\langle\langle \hk_1, \hepsilon_{1}|  
~V(x=1; \hk_2, \hepsilon_2)
\alpha^{\mu}_{-1}|\hqs\rangle
&=
~\epsilon^{\mu_{1}}_{a_{1}}\epsilon^{\mu_{2}}_{a_{2}}
V^{(1 2 3)}
_{\mu_{1};~\mu_{2};~\mu}(k_{1}, k_{2}, \qs) 
\nonumber\\
\sqrt{ -\cCz \cNz^4 }
\langle\langle \hqs| \alpha^{\nu}_{1}
~V(x=1; \hk_3, \hepsilon_4)
~ |\hk_4, \hepsilon_{4}\rangle
&=
V^{(1 2 3)}
_{\nu;~\mu_{3};~\mu_{4}}(-\qs, k_{3}, k_{4})
~\epsilon^{\mu_{3}}_{a_{3}}\epsilon^{\mu_{4}}_{a_{4}}
.
\end{align}
As discussed in the previous section this is not possible since the
string truncated Green function 
is not cyclically invariant while the \subvertex{} is,
the proper expressions are 
\COMMENTOBUH{ terms proportional to \(k_{1}\cdot k_{2}\)}
\begin{align}
\sqrt{-\cCz \cNz^4 }
\langle\langle \hk_1, \hepsilon_{1}|  
~V(x=1; \hk_2, \hepsilon_2)
\alpha_{\mu\, -1}|\hqs\rangle
&=
~\epsilon^{\mu_{1}}_{a_{1}}\epsilon^{\mu_{2}}_{a_{2}}
V^{(1 2 3)}
_{\mu_{1};~\mu_{2};~\mu}(k_{1}, k_{2}, \qs) 
+ O(\qs_\mu) 
+ O(\qs^{2}) 
\nonumber\\
\sqrt{-\cCz \cNz^4 }
\langle\langle \hqs| \alpha_{\nu 1}
~V(x=1; k_3, \epsilon_4)
~ |k_4, \epsilon_{4}\rangle
&=
V^{(1 2 3)}
_{\nu;~\mu_{3};~\mu_{4}}(-q, k_{3}, k_{4})
~\epsilon^{\mu_{3}}_{a_{3}}\epsilon^{\mu_{4}}_{a_{4}}
+ O(\qs_\mu)
+ O(\qs^{2}) 
,
\end{align}
where the terms in $V^{(1 2 3)}$ proportional to $k_1 \cdot k_2$ contribute as
$\qs^2$ because of momentum conservation.
It is possible to use the previous less restrictive identification
since  for example
$\qs^\mu V^{(1 2 3)}
_{\nu;~\mu_{3};~\mu_{4}}(-\qs, k_{3}, k_{4}) 
~\epsilon^{\mu_{3}}_{a_{3}}\epsilon^{\mu_{4}}_{a_{4}}
 \propto \qs^2$ 
so that the propagator pole is canceled.
This happens because 
$\qs^\mu V^{(1 2 3)}
_{\nu;~\mu_{3};~\mu_{4}}=0$
when we take $\epsilon=\qs$ and the gluon physical, 
i.e. $q^2=0$ because of gauge invariance.

Finally we get the constraints
\footnote{
From these equations and eq. (\ref{eq:c1,3 and C0}) it follows that
\(\cCz \cNz^{2}= \tap^{\oh D -1} \) and then
\(\cNz=\frac{g}{\ap}\) and \(\cCz= 1/(2 g)^{2} \tap^{\oh D +1}\).}
\begin{align}
\label{c_constraints}
c_1 - c_2
= 
\frac{
3 c_3 - 3 c_{4} + 3c_5 - 3 c_6
}{\tap}
=
- \ii \cCz \cNz^3 \tap^{2-\oh D}
=
 \sqrt{- \cCz \cNz^4 } \tap^{2-\oh D}
.
\end{align}
No constraints are obtained on the other coefficients since all of
them contribute terms proportional to $\qs^2$.

It is also interesting and consistent with the previous line of
thought to consider what happens when the gluons \(1\) and \(4\) are
not physical. 
In this case the difference in eq. (\ref{eq:s_ch_subtr_semplified})
must be a sum of terms proportional to one of the following factors
\(\hk_{1}^{2}\), \(\hk_{4}^{2}\), \(\he_{1}\cdot\hk_{1}\) or
\(\he_{4}\cdot\hk_{4}\) since these are  vanishing when the particles
are physical.
A direct computation reveals that all of these terms are actually
present.
This means that the string truncated partially off shell \(N=4\) Green
function when subtracted the Feynman diagrams still has poles.
This seems wrong but it is not so.
The reason is that using the naive factorization we cannot compare
directly the truncated Green functions since  the \(N=3\) truncated
Green functions do not match perfectly between string theory
and  the usual EFT.
Nevertheless the \(S\) matrix elements must match and not only the
full \(S\) matrix but also the color ordered sub-pieces.

\subsection{Computing the contact terms up to \(k^{2}\) order.}
\label{sec:N4_old}
In order to compute the \(N=4 \) \subvertices{}
we need to compute the usual string amplitude 
and then expand in momentum powers. 
We write the basic amplitude as
\begin{eqnarray}
  \label{eq:N4_computation0}
A_{1 2 3 4}&=
\int_{0}^{1} d y~
\langle\langle \hk_1, \hepsilon_{1}|  
~V(x=1; \hk_2, \hepsilon_2)
~y^{L_{0}-2}
~V(x=1; \hk_3, \hepsilon_4)
~ |\hk_4, \hepsilon_{4}\rangle
  \nonumber\\
&=
\int_{0}^{1} d y~
\langle\langle \hk_1, \hepsilon_{1}|  
~V(x=1; \hk_2, \hepsilon_2)
~y^{L_{0}-2}
~V(x=1; \hk_3, \hepsilon_4)
~ |\hk_4, \hepsilon_{4}\rangle
  \nonumber\\
&=
\int_{0}^{1} d y~
\langle\langle \hk_1, \hepsilon_{1}|  
~V(1; \hk_2, \hepsilon_2)
~y^{(\hk_{3}^{2}+1)-2}
~V(y; \hk_3, \hepsilon_4)
~ y^{L_{0}}|\hk_4, \hepsilon_{4}\rangle
\end{eqnarray}
The explicit expression for this contribution to the amplitude is given
\begin{align}
A_{1234}
=&
\mymeno\Big[
\left(1- \oh \hs -\oh \hu\right) C_{(0,0)}
+ \frac{1}{-\hs/2} C_{(1,0)}
+ \frac{1}{-\hu/2} C_{(0,1)}
\nonumber\\
&- \left(1- \frac{\hu}{\hs}\right) \frac{1}{1-\hs/2} C_{(2,0)}
- \left(1- \frac{\hs}{\hu}\right) \frac{1}{1-\hu/2} C_{(0,2)}
\Big]
\nonumber\\
&\times
\frac{\Gamma\left(1-\oh \hs\right)\, \Gamma\left(1-\oh \hu\right)}
{\Gamma\left(1-\oh \hs-\oh \hu\right)}
\delta_{\sum \hk}
,
\end{align}
where the coefficients $C_{(\cdot, \cdot)}$ are given in eq.s
(\ref{C00},\ref{C01},\ref{C10},\ref{C20},\ref{C02}) in appendix
\ref{sec:details_4_gluons_correlator}. 
In order to compare with the EFT we need to expand the previous
expression in momentum powers, explicitly we get
\begin{align}
A_{1234}
=&
\mymeno\Big\{
%
&+&\Big[
-C_{(2,0)}|_{k^0} \frac{u}{s}
-C_{(0,2)}|_{k^0} \frac{s}{u}
+C_{(1,0)}|_{k^2} \frac{1}{-s/2}
+C_{(0,1)}|_{k^2} \frac{1}{-u/2}
\nonumber\\
& &&
-C_{(2,0)}|_{k^0} -C_{(0,2)}|_{k^0} +C_{(0,0)}|_{k^0}
\Big]
\nonumber\\
%
& &+&\Big[
-C_{(2,0)}|_{k^2} \frac{u}{s}
-C_{(0,2)}|_{k^2} \frac{s}{u}
+C_{(1,0)}|_{k^4} \frac{1}{-s/2}
+C_{(0,1)}|_{k^4} \frac{1}{-u/2}
\nonumber\\
& &&
+\left( -C_{(2,0)}|_{k^0} -C_{(0,2)}|_{k^0} +C_{(0,0)}|_{k^0} \right)
     \left(-\oh s-\oh u\right)
\nonumber\\
& &&
-C_{(2,0)}|_{k^2} -C_{(0,2)}|_{k^2} +C_{(0,0)}|_{k^2}
\Big]
\nonumber\\
%
& &+&\Big[
-C_{(2,0)}|_{k^4} \frac{u}{s}
-C_{(0,2)}|_{k^4} \frac{s}{u}
\nonumber\\
& &&
-C_{(2,0)}|_{k^4} -C_{(0,2)}|_{k^4} +C_{(0,0)}|_{k^4}
\nonumber\\
& &&
+\left( -C_{(2,0)}|_{k^2} -C_{(0,2)}|_{k^2} +C_{(0,0)}|_{k^2}
+(\Gamma'(1)^2-\Gamma''(1)) C_{(0,1)}|_{k^2}
 \right) \frac{-s}{2}
\nonumber\\
& &&
+\left( -C_{(2,0)}|_{k^2} -C_{(0,2)}|_{k^2} +C_{(0,0)}|_{k^2}
+(\Gamma'(1)^2-\Gamma''(1)) C_{(1,0)}|_{k^2}
 \right) \frac{-u}{2}
\nonumber\\
& &&
-\left( C_{(2,0)}|_{k^0} +(\Gamma'(1)^2-\Gamma''(1)) C_{(0,2)}|_{k^0} \right) 
\left(\frac{-s}{2}\right)^2
\nonumber\\
& &&
-\left( C_{(0,2)}|_{k^0} +(\Gamma'(1)^2-\Gamma''(1)) C_{(2,0)}|_{k^0} \right) 
\left(\frac{-u}{2}\right)^2
\nonumber\\
& &&
-\left( 
(\Gamma'(1)^2-\Gamma''(1)+1) 
\left( -C_{(2,0)}|_{k^0} -C_{(0,2)}|_{k^0} +C_{(0,0)}|_{k^0} \right)
-C_{(0,0)}
\right) 
\frac{-s}{2} \frac{-u}{2}
\Big]
\nonumber\\
%
& &+&
O(k^6)
\Big\}
\delta_{\sum \hk}
,
\end{align}
where $C_{(\cdot, \cdot)}|_{k^n}$ stands for the part with \(n\) momentum
powers in the coefficient $C_{(\cdot, \cdot)}$.

Since now we are considering the string amplitude for all possible
values of the momenta we must
subtract all the Feynman diagrams built with \subvertices{} which have
the proper color ordering and poles in the same channels of the string
amplitude, both \(s\) and \(u\) for the amplitude \(A_{1 2 3 4}\).  
When canceling the poles we get again eq.s (\ref{c_constraints}).
The explicit computation gives at \(k^{0}\) order
\COMMENTOBUH{should be $c_{1}-c_{2}$, I should have changed
  \(c_{2,4}\rightarrow - c_{2,4}\) so i do it the I should check}
\begin{align}
\label{eq:subvertex_N4_k0}
\ap &\cCz \cNz^4 A_{1 2 3 4}|_{k^0} 
&-&
\epsilon^{\mu_{1}}_{a_{1}}\epsilon^{\mu_{2}}_{a_{2}}
V^{(1 2 3)}
_{\mu_{1};~\mu_{2};~\mu}(k_{1}, k_{2}, \qs)|_{k^1}
\frac{\delta^{\mu\nu} }{\qs^2}
V^{(1 2 3)}
_{\nu;~\mu_{3};~\mu_{4}}(-\qs, k_{3}, k_{4})|_{k^1}
~\epsilon^{\mu_{3}}_{a_{3}}\epsilon^{\mu_{4}}_{a_{4}}
\delta_{\sum k_{i}}
\nonumber\\
&
&-&
~\epsilon^{\mu_{4}}_{a_{4}}\epsilon^{\mu_{1}}_{a_{1}}
V^{(1 2 3)}
_{\mu_{4};~\mu_{1};~\mu}(k_{4}, k_{1}, \qs)|_{k^1}
\frac{\delta^{\mu\nu} }{\qs^2}
V^{(1 2 3)}
_{\nu;~\mu_{2};~\mu_{3}}(-\qs, k_{2}, k_{3})|_{k^1}
 ~\epsilon^{\mu_{2}}_{a_{2}}\epsilon^{\mu_{3}}_{a_{3}}
\delta_{\sum k_{i}}
\nonumber\\
&&=&
\Bigl[
-c_1\, c_2\,
\left(
\epsilon_1\cdot \epsilon_2\,
\epsilon_3\cdot \epsilon_4\,
+
\epsilon_1\cdot \epsilon_4\,
\epsilon_2\cdot \epsilon_3\,
\right)
\nonumber\\
&&&
-\oh (c_1 -c_2)^2
\epsilon_1\cdot \epsilon_3\,
\epsilon_2\cdot \epsilon_4
\Bigr]
\delta_{\sum k_{i}}
,
\end{align}
along with the constraint from pole cancellation
\begin{equation}
\label{eq:constraint_N4_k0}
(c_{1} - c_{2} )^{2} = - \cCz \cNz^{4} \tap^{3-\oh D}
.
\end{equation}
Notice that the previous expression is cyclically invariant therefore
we can interpret it as the quartic \subvertex{} at order \(k^{0}\).

From this expression it is then clear that the choice \(c_{2}=0\) (or
\(c_{1}=0\)) is the most economical. This is exactly the choice
suggested by the string.

The previous \subvertex{} at order $k^0$ then becomes in the gauge
suggested by the string
\begin{align}
V&^{(1234)}_{1234}(k_1,k_2,k_3,k_4) |_{k^0}
=
+2 g^2
\bigg\{
\epsilon_1 \cdot \epsilon_3\, \epsilon_2\epsilon_4
\bigg\}
\delta_{\sum k_{i}}
,
\end{align}
which is the  \subvertex{} depicted in figure
\ref{fig:Color_Ordered_Feynman_Rules}. 

\COMMENTOBUH{
It is exactly this \subvertex{} which is pictured in  figure \ref{fig:Color_Ordered_Feynman_Rules}.
}
The quartic vertex at \(k^{0}\) order reads in general
\begin{align}
V_{[4]}
&
(k_1,k_2,k_3,k_4) |_{k^0}
=
\nonumber\\
%
\phantom{V_{[4]}}
&
\Bigl\{
+\epsilon_{a_1}\cdot\epsilon_{a_2}\,
\epsilon_{a_3}\cdot\epsilon_{a_4}\,
\Big[
-
c_1\,c_2\,
tr\left(
   \left\{ T_{a_1}, T_{a_2}\right\}
   \left\{ T_{a_3}, T_{a_4}\right\}
\right)
\nonumber\\
\phantom{V_{[4]}}
&
\phantom{\Bigl\{
+\epsilon_{a_1}\cdot\epsilon_{a_2}\,
\epsilon_{a_3}\cdot\epsilon_{a_4}\,
\Big[
} 
-
\oh (c_1-c_2)^2
tr\left( 
   T_{a_1}\, T_{a_4}\, T_{a_2}, T_{a_3}
   + T_{a_1}\, T_{a_3}\, T_{a_2}, T_{a_4}
\right)
\Big]
\nonumber\\
%
\phantom{V_{[4]}}
&
+
\epsilon_{a_1}\cdot\epsilon_{a_3}\,
\epsilon_{a_2}\cdot\epsilon_{a_4}\,
\Big[
-
c_1\,c_2\,
tr\left(
   \left\{ T_{a_1}, T_{a_3}\right\}
   \left\{ T_{a_2}, T_{a_4}\right\}
\right)
\nonumber\\
\phantom{V_{[4]}}
&
\phantom{\Bigl\{
+\epsilon_{a_1}\cdot\epsilon_{a_2}\,
\epsilon_{a_3}\cdot\epsilon_{a_4}\,
\Big[
} 
-
\oh (c_1-c_2)^2
tr\left( 
   T_{a_1}\, T_{a_4}\, T_{a_3}, T_{a_2}
   + T_{a_1}\, T_{a_2}\, T_{a_2}, T_{a_4}
\right)
\Big]
\nonumber\\
%
\phantom{V_{[4]}}
&
+\epsilon_{a_1}\cdot\epsilon_{a_4}\,
\epsilon_{a_2}\cdot\epsilon_{a_3}\,
\Big[
-
c_1\,c_2\,
tr\left(
   \left\{ T_{a_1}, T_{a_4}\right\}
   \left\{ T_{a_2}, T_{a_3}\right\}
\right)
\nonumber\\
\phantom{V_{[4]}}
&
-
\phantom{\Bigl\{
+\epsilon_{a_1}\cdot\epsilon_{a_2}\,
\epsilon_{a_3}\cdot\epsilon_{a_4}\,
\Big[
} 
\oh (c_1-c_2)^2
tr\left( 
   T_{a_1}\, T_{a_3}\, T_{a_4}, T_{a_2}
   + T_{a_1}\, T_{a_2}\, T_{a_4}, T_{a_3}
\right)
\Big]
\Bigl\}
\delta_{\sum k_{i}}
.
\end{align}
The explicit computation at $k^2$ order requires
\begin{align}
(c_1 - c_2) ( -3 c_3 + 3 c_4 + 3 c_5 -  3 c_6 )
=
\cCz \cNz^4
\tap^{4-\oh D} 
\end{align}
because of pole cancellation and gives 
\begin{align}
\label{eq:general_v4_k2}
\ap \cCz \cNz^4 A_{1 2 3 4}|_{k^2} 
&-
\frac{\delta^{\mu\nu} }{\qs^2}
\left.\left[
~\epsilon^{\mu_{1}}_{a_{1}}\epsilon^{\mu_{2}}_{a_{2}}
V^{(1 2 3)}
_{\mu_{1};~\mu_{2};~\mu}(k_{1}, k_{2}, \qs)
V^{(1 2 3)}
_{\nu;~\mu_{3};~\mu_{4}}(-\qs, k_{3}, k_{4})
~\epsilon^{\mu_{3}}_{a_{3}}\epsilon^{\mu_{4}}_{a_{4}}
\right]\right|_{k^4}
\delta_{\sum k_{i}}
\nonumber\\
&-
\frac{\kappa\,\delta^{\mu\nu} }{\qs^2}
\left.\left[
~\epsilon^{\mu_{4}}_{a_{4}}\epsilon^{\mu_{1}}_{a_{1}}
V^{(1 2 3)}
_{\mu_{4};~\mu_{1};~\mu}(k_{4}, k_{1}, \qs)
V^{(1 2 3)}
_{\nu;~\mu_{2};~\mu_{3}}(-\qs, k_{2}, k_{3})
 ~\epsilon^{\mu_{2}}_{a_{2}}\epsilon^{\mu_{3}}_{a_{3}}
\right]\right|_{k^4}
\delta_{\sum k_{i}}
\nonumber\\
%
%
=&
\Bigg\{
-\oh [ c_1 ( - 3 c_4- c_5 + 2 c_6) + c_2 ( -3 c_3 + 2 c_5 - c_6) ]
\nonumber\\
&
\phantom{\Bigg\{-\oh}
\times
[
\epsilon_1 \cdot \epsilon_2\, \epsilon_3\cdot k_4\, \epsilon_4\cdot  k_3
+
\mbox{3 terms obtained by cycling $(1234)$ in the previous term}
]
\nonumber\\
\nonumber\\
%
%
%
%
+
\bigg\{
&
+\oh \epsilon_1 \cdot \epsilon_2\, \epsilon_3\cdot \epsilon_4
\Big[ 
\left( \oh \tap^{4-\oh D} \cCz \cNz^4 \right) s
+\left( \oh \tap^{4-\oh D} \cCz \cNz^4  +c_1 c_9+ c_2 c_{12}\right) u
\nonumber\\
&\phantom{+ +\oh \epsilon_1 \cdot \epsilon_2\, \epsilon_3\epsilon_4\Big[ }
+\left( \oh \tap^{4-\oh D} \cCz \cNz^4  +c_1 c_{12}+ c_2 c_{9}\right) t
\Big] 
\nonumber\\
&+
\mbox{1 term obtained by cycling $(1234)$ in the previous term}
\bigg\}
\nonumber\\
-
&
\oh \epsilon_3 \cdot \epsilon_1\, \epsilon_2\cdot \epsilon_4
\left[ 
\left( \oh \tap^{4-\oh D} \cCz \cNz^4 \right) u
+\left( \oh \tap^{4-\oh D} \cCz \cNz^4 \right) s
\right]
\nonumber\\
%
%
+\bigg\{
+&\epsilon_1 \cdot \epsilon_2\, \epsilon_3\cdot k_1\, \epsilon_4\cdot  k_1
\oh
\Big[
 \tap^{4-\oh D} \cCz \cNz^4
+ c_1 ( -3 c_4 -c_5+2 c_6)
\nonumber\\
&\phantom{\epsilon_1 \cdot \epsilon_2\, \epsilon_3\cdot k_1\,
  \epsilon_4\cdot  k_1 \oh } 
+ c_2 ( -3 c_3 + 2c_5-c_6-2c_7+2c_8+2c_{10}-2c_{11})
\Big]
\nonumber\\
+&\epsilon_1 \cdot \epsilon_2\, \epsilon_3\cdot k_1\, \epsilon_4\cdot  k_2
\oh
\Big[
 \tap^{4-\oh D} \cCz \cNz^4
+ c_1 ( -3c_4 - c_5 +2c_6 +2c_8 -2c_{11})
\nonumber\\
&\phantom{\epsilon_1 \cdot \epsilon_2\, \epsilon_3\cdot k_1\,
  \epsilon_4\cdot  k_1 \oh } 
+ c_2 ( -3c_3 +2c_5 -c_6 -2c_7 +2c_{10})
\Big]
\nonumber\\
+&\epsilon_1 \cdot \epsilon_2\, \epsilon_3\cdot k_1\, \epsilon_4\cdot  k_2
\oh
\Big[
 \tap^{4-\oh D} \cCz \cNz^4
\nonumber\\
&\phantom{\epsilon_1 \cdot \epsilon_2\, \epsilon_3\cdot k_1\,
  \epsilon_4\cdot  k_1 \oh } 
+ c_1 ( -3c_4 - c_5 +2c_6 -4c_7 -2c_8 +4c_9 +2c_{10} +2c_{11} -2c_{12})
\nonumber\\
&\phantom{\epsilon_1 \cdot \epsilon_2\, \epsilon_3\cdot k_1\,
  \epsilon_4\cdot  k_1 \oh } 
+ c_2 ( -3c_3 +2c_5 -c_6 +2c_7 +4c_8 -2c_9 -2c_{10} -4c_{11} +4c_{12})
\Big]
\nonumber\\
+&\epsilon_1 \cdot \epsilon_2\, \epsilon_3\cdot k_2\, \epsilon_4\cdot  k_2
\oh
\Big[
 \tap^{4-\oh D} \cCz \cNz^4
+ c_1 ( -3c_4 - c_5 +2c_6 -2c_7 +2c_8 )
\nonumber\\
&\phantom{\epsilon_1 \cdot \epsilon_2\, \epsilon_3\cdot k_1\,
  \epsilon_4\cdot  k_1 \oh } 
+ c_2 ( -3c_3 +2c_5 -c_6 )
\Big]
\nonumber\\
&+
\mbox{3*4 terms obtained by cycling $(1234)$ in the previous 4 terms}
\bigg\}
\nonumber\\
+\bigg\{
+&\epsilon_1 \cdot \epsilon_3\, \epsilon_2\cdot k_1\, \epsilon_4\cdot  k_1
\Big [-\oh  \tap^{4-\oh D} \cCz \cNz^4 \Big ]
\nonumber\\
+&\epsilon_1 \cdot \epsilon_3\, \epsilon_2\cdot k_1\, \epsilon_4\cdot  k_3
\Big [-\oh  \tap^{4-\oh D} \cCz \cNz^4 
-(c_1-c_2)( -2c_7 +c_9 +2c_{10} -c_{12}) \Big ]
\nonumber\\
+&\epsilon_1 \cdot \epsilon_3\, \epsilon_2\cdot k_3\, \epsilon_4\cdot  k_1
\Big [-\oh  \tap^{4-\oh D} \cCz \cNz^4 
+(c_1-c_2)( +2c_7 -c_9 -2c_{11} +c_{12}) ]
\Big ]
\nonumber\\
+&\epsilon_1 \cdot \epsilon_3\, \epsilon_2\cdot k_3\, \epsilon_4\cdot  k_3
\Big [-\oh  \tap^{4-\oh D} \cCz \cNz^4 \Big ]
\nonumber\\
&+
\mbox{4 terms obtained by cycling $(1234)$ in the previous 4 terms}
\bigg\}
\Bigg\}
\delta_{\sum k_{i}}
.
\end{align}
Again the previous result can be interpreted as the $N=4$ gluon \subvertex{}
since it is cyclically invariant.

Moreover the suggestion of string theory is the more economical 
since all terms coming from \subvertices{} vanish when only $c_1$ and
$c_3$ are different from zero. 
This can also be understood by the fact that the $3$ gluon
\subvertex{} suggested by the string has the minimal contain to cancel
the poles in the $4$ gluons amplitude.

The previous \subvertex{} at order $k^2$ 
becomes in the gauge suggested by the string
\begin{align}
V&^{(1234)}_{1234}(k_1,k_2,k_3,k_4) |_{k^2}
=
-4 \ap g^2
\bigg\{
\nonumber\\
&+
\Big[
\epsilon_1 \cdot \epsilon_2\, \epsilon_3\epsilon_4
+
\mbox{1 term from cycling $(1234)$}
\Big]
\frac{t}{2}
\nonumber\\
&
+
\Big[
\epsilon_1 \cdot \epsilon_3\, \epsilon_2\cdot\epsilon_4
\Big]
\frac{t}{2}
\nonumber\\
&
-
\Big[
+\epsilon_1 \cdot \epsilon_2\, \epsilon_3\cdot k_4\, \epsilon_4\cdot  k_3
+
\mbox{4 terms from cycling $(1234)$}
\Big]
\nonumber\\
&
+
\Big[
+\epsilon_1 \cdot \epsilon_3\, \epsilon_2\cdot k_4\, \epsilon_4\cdot  k_2
+
\mbox{1 term from cycling $(1234)$}
\Big]
\bigg\}
\delta_{\sum k_{i}}
.
\end{align}
Finally the full the quartic vertex at \(k^{2}\) order in string gauge
reads
\begin{align}
V_{[4]} |_{k^2}
=&
\nonumber\\
%
\bigg\{
&+\
V^{(1234)}_{a_1 a_2 a_3 a_4}\,
tr\left( 
   T_{a_1}\, T_{a_4}\, T_{a_2}, T_{a_3}
   + T_{a_1}\, T_{a_3}\, T_{a_2}, T_{a_4}
\right)
\nonumber\\
&+
V^{(1234)}_{a_1 a_4 a_2 a_3}\,
tr\left( 
   T_{a_1}\, T_{a_4}\, T_{a_3}, T_{a_2}
   + T_{a_1}\, T_{a_2}\, T_{a_2}, T_{a_4}
\right)
\nonumber\\
&+
V^{(1234)}_{a_1 a_3 a_4 a_2}\,
tr\left( 
   T_{a_1}\, T_{a_3}\, T_{a_4}, T_{a_2}
   + T_{a_1}\, T_{a_2}\, T_{a_4}, T_{a_3}
\right)
\bigg\}
\delta_{\sum k_{i}}
.
\end{align}
because $V^{(1234)}_{1234}=V^{(1234)}_{1432}$.
We have also substituted $\epsilon_i \rightarrow \epsilon_{a_i}$.
This expression is by far simpler than the one obtained in the usual
Feynman gauge where a lot of $c$.s are different from zero in
eq. (\ref{eq:general_v4_k2}). 

\vskip 1cm
\noindent {\large {\bf Acknowledgments}}

This work is partially supported by the Compagnia di San Paolo
contract ``MAST: Modern Applications of String Theory''
TO-Call3-2012-0088 and by the MIUR PRIN Contract 2015MP2CX4 
``Non-perturbative Aspects Of Gauge Theories And Strings"

\appendix

\section{Conventions}
\label{app:conventions}
We write the open string expansion for the dimensionless field
$\hX=\sqtap X$ as
\begin{equation}
  \label{eq:X_exp}
  \hX^\mu(u,\bu)
=\oh( \hX_L^\mu(u)+ \hX_R^\mu(\bu))
\end{equation}
with
$$u=e^{\tau_E+i \ii\sigma}\in \Hp$$
and
\begin{eqnarray}
  \label{eq:XL-XR-exp}
  \hX_L(u)&=
\hx_0 +\hy_0
-\ii \alpha_0~\ln(u)
+\ii  \sum_{n\ne0} \frac{\alpha_n}{n} u^{-n}
\nonumber\\
  \hX_R(\bu)&=
\hx_0 -\hy_0
-\ii \alpha_0~\ln(\bu)
+\ii \sum_{n\ne0} \frac{\alpha_n}{n} \bu^{-n}
\end{eqnarray}

The commutation relations read
\begin{equation}
  \label{eq:comm_rel}
  [\alpha^\mu_n,~ \alpha^\nu_m]= n~ \delta^{\mu\nu}~ \delta_{m+n,0}
.
\end{equation}

The mass shell condition reads
\begin{equation}
  \label{eq:L0}
  L_0^{(X)} |phys\rangle 
= ( \oh \alpha_0^2+ \sum_{n=1}^\infty \alpha_{-n}\cdot\alpha_n ) |phys\rangle
= | phys\rangle
.
\end{equation}

The momentum states are defined as
\begin{equation}
e^{i \hk \cdot \hx_{0}} |0\rangle = |k\rangle,~~~~
\langle 0 |e^{-i \hk \cdot \hx_{0}}  = \langle\langle k|
.
\end{equation}

\COMMENTOBUH{
\begin{eqnarray}
  \label{eq:vertex_pieces}
  :\ee^{i \hk \hX(x,x)}:
&=
\ee^{i \hk \hx_0}~
\ee^{+\hk \sum_{n=1}^\infty \frac{\alpha_{-n}}{n} x^n}~
|x|^{\hk \alpha_0}~
\ee^{-\hk \sum_{n=1}^\infty \frac{\alpha_{n}}{n} x^{-n}}
\nonumber\\
\du \hX(u, \bu)
&=
\oh \du \hX_L(u)
=
-\ii \sum_{n} \frac{\alpha_n}{u^{n+1}}
\end{eqnarray}
}

\section{YM conventions}
\label{sec:YMconv}
The Euclidean YM Lagrangian reads
\begin{equation}
{\cal L}_{E}
= +\frac{1}{4 \kappa} tr( F_{\mu\nu} F_{\mu\nu} )
= +\frac{1}{4 }  F^a_{\mu\nu} F^a_{\mu\nu},
\end{equation}
since we normalize the generators $T^a=T^{a\dagger}$ as
\begin{equation}
\label{eq:tr_norm}
tr(T_a T_b)= \kappa \delta_{a b},
~~~
[T_a, T_b]= i f_{a b c } T^c.
\end{equation}
It then follows that $tr(T_a [T_b, T_c]) = i \kappa f_{a b c }$.
We define the field strength of the gauge field 
$A= A_\mu d x^\mu = A_\mu^a T^a d x^\mu$ 
as
\begin{align}
F &= d A - i g A\wedge A = \oh F_{\mu\nu} d x^\mu \wedge d x^\nu
\nonumber\\
 F_{\mu\nu} &= \partial_\mu A_\nu -  \partial_\nu A_\mu
-i g [  A_\mu,  A_\nu]
\nonumber\\
F^a_{\mu\nu} &= \partial_\mu A^a_\nu -  \partial_\nu A^a_\mu
+ g f_{a b c }  A^b_\mu  A^c_\nu
.
\end{align}
The Lagrangian then becomes
\COMMENTOBUH{ Changed signs passing to Euclidean}
\begin{align}
{\cal L}_{E}
=&
\frac{1}{2 \kappa} tr\Big( \partial_\mu A_\nu 
(\partial_\mu A_\nu -  \partial_\nu A_\mu) \Big)
\nonumber\\
&
+
\frac{-i g}{\kappa} tr \Big( \partial_\mu A_\nu [ A_\mu,  A_\nu] \Big)
\nonumber\\
&+\frac{(-i g)^2}{4 \kappa } 
tr \Big(  [A_\mu,  A_\nu]  [A_\mu,  A_\nu] \Big)
.
\end{align}
\COMMENTOBUH{
\begin{align}
{\cal L}_{E}
=&
+\oh \partial_\mu A^a_\nu (\partial_\mu A^a_\nu -  \partial_\nu
A^a_\mu)
\nonumber\\
&
+g f_{a b c } \partial_\mu A^a_\nu  A^b_\mu  A^c_\nu
\nonumber\\
&+\frac{1}{4} g^2 f_{a b c } f_{a d e }
 A^b_\mu  A^c_\nu  A^d_\mu  A^e_\nu
.
\end{align}
}
Let us rewrite the cubic interaction term in momentum space
\begin{align}
-S_{E\,[3]}
=&
-\int \prod_{i=1}^3 \frac{d^D k_i}{(2\pi)^D}
(2\pi)^D \delta^D(k_1+k_2+k_3)
\times
\frac{i g}{\kappa} tr(T_{a_1}[T_{a_2}, T_{a_3}]) 
(i k_{1 \mu_2})  \delta_{\mu_2 \mu_3}
\nonumber\\
&
\phantom{-\int \prod_{i=1}^3 \frac{d^D k_i}{(2\pi)^D}
(2\pi)^D \delta^D(k_1+k_2+k_3)
}
\times
\epsilon_{a_1 \mu_1}(k_1)\epsilon_{a_2 \mu_2}(k_2)\epsilon_{a_3 \mu_3}(k_3)
\nonumber\\
=&
\int \prod_{i=1}^3 \frac{d^D k_i}{(2\pi)^D}
\delta_{\sum k}
\times
\frac{1}{3!} 
\frac{- g}{\kappa} tr(T_{a_1}[T_{a_2}, T_{a_3}]) 
\nonumber\\
&\times
[
(k_{1 \mu_2} - k_{3 \mu_2}) \delta_{\mu_3 \mu_1}
+
(k_{3 \mu_1} - k_{2 \mu_1}) \delta_{\mu_2 \mu_3}
+
(k_{2 \mu_3} - k_{1 \mu_3}) \delta_{\mu_1 \mu_2}
]
\nonumber\\
&\times
\epsilon_{a_1 \mu_1}(k_1)\epsilon_{a_2 \mu_2}(k_2)\epsilon_{a_3 \mu_3}(k_3)
,
\end{align}
then it follows that
\begin{align}
V_{a_1 \mu_1; a_2 \mu_2; a_3 \mu_3}(k_1, k_2, k_3)
=&
V^{(1 2 3)}_{a_1 \mu_1; a_2 \mu_2; a_3 \mu_3}(k_1, k_2, k_3)
+V^{(1 2 3)}_{a_1 \mu_1;  a_3 \mu_3; a_2 \mu_2}(k_1, k_3, k_2)
\nonumber\\
&=
\frac{- g}{\kappa} tr(T_{a_1}[T_{a_2}, T_{a_3}])
\nonumber\\
&\times
[
(k_{1 \mu_2} - k_{3 \mu_2}) \delta_{\mu_3 \mu_1}
+
(k_{3 \mu_1} - k_{2 \mu_1}) \delta_{\mu_2 \mu_3}
+
(k_{2 \mu_3} - k_{1 \mu_3}) \delta_{\mu_1 \mu_2}
]
\delta_{\sum k}
,
\end{align}
and
\begin{align}
V^{(1 2 3)}_{a_1 \mu_1; a_2 \mu_2; a_3 \mu_3}(k_1, k_2, k_3)
&=
\frac{- g}{\kappa} tr(T_{a_1} T_{a_2} T_{a_3})
\nonumber\\
&\times
[
(k_{1 \mu_2} - k_{3 \mu_2}) \delta_{\mu_3 \mu_1}
+
(k_{3 \mu_1} - k_{2 \mu_1}) \delta_{\mu_2 \mu_3}
+
(k_{2 \mu_3} - k_{1 \mu_3}) \delta_{\mu_1 \mu_2}
]
\delta_{\sum k}
.
\end{align}

Similarly we write the quartic action as
\begin{align}
-S_{E\,[4]}
=&
-\int \prod_{i=1}^4 \frac{d^D k_i}{(2\pi)^D}
(2\pi)^D \delta^D(k_1+k_2+k_3+k_3)
\times
\frac{(-i g)^2}{4\kappa} 
tr([T_{a_1}, T_{a_2}] [T_{a_3}, T_{a_4}]) 
 \delta_{\mu_1 [ \mu_3}  \delta_{\mu_4 ] \mu_2}
\nonumber\\
&
\phantom{-\int \prod_{i=1}^3 \frac{d^D k_i}{(2\pi)^D}
(2\pi)^D \delta^D(k_1+k_2+k_3)
}
\times
\epsilon_{a_1 \mu_1}(k_1)\epsilon_{a_2 \mu_2}(k_2)\epsilon_{a_3 \mu_3}(k_3)
\epsilon_{a_4 \mu_4}(k_4)
,
\end{align}
the using the symmetries $1\leftrightarrow 2$, $3\leftrightarrow 4$
and $(1,2)\leftrightarrow (3,4)$ we sum over the remaining $4!/2^3=3$.
Using the previous symmetries we can always set $1$ in the first place
of the permutation and then we are left with $ 1 2 3 4$, $1 3 4 2$ and
$1 4 2 3$. So we get
\begin{align}
V&_{a_1 \mu_1; a_2 \mu_2; a_3 \mu_3; a_4 \mu_4}(k_1, k_2, k_3, k_4)
=
\nonumber\\
=&
V^{(1 2 3 4)}_{a_1 \mu_1; a_2 \mu_2; a_3 \mu_3; a_4 \mu_4}(k_1, k_2, k_3, k_4)
+V^{(1 2 3 4)}_{a_1 \mu_1;  a_3 \mu_3; a_4 \mu_4; a_2 \mu_2}(k_1, k_3, k_4, k_2)
+V^{(1 2 3 4)}_{a_1 \mu_1;  a_4 \mu_4; a_2 \mu_2; a_3 \mu_3}(k_1, k_4, k_2, k_3)
\nonumber\\
=&
\frac{2 g^2}{\kappa} 
\Big[ 
tr([T_{a_1}, T_{a_2}] [T_{a_3}, T_{a_4}]) 
 \delta_{\mu_1 [ \mu_3}  \delta_{\mu_4 ] \mu_2}
+
tr([T_{a_1}, T_{a_3}] [T_{a_4}, T_{a_2}]) 
 \delta_{\mu_1 [ \mu_4}  \delta_{\mu_2 ] \mu_3}
\nonumber\\
&
\phantom{\frac{8 g^4}{\kappa} \Big[ }
+
tr([T_{a_1}, T_{a_4}] [T_{a_2}, T_{a_3}]) 
 \delta_{\mu_1 [ \mu_2}  \delta_{\mu_3] \mu_4}
\Big]
\delta_{\sum k}
,
\end{align}
from which it follows
\begin{align}
V&^{(1 2 3 4)}_{a_1 \mu_1; a_2 \mu_2; a_3 \mu_3; a_4 \mu_4}(k_1, k_2,
  k_3, k_4)
=
\nonumber\\
&=
\frac{ g^2}{\kappa} 
tr(T_{a_1} T_{a_2} T_{a_3} T_{a_4}) 
\Big[ 
2 \delta_{\mu_1  \mu_3}  \delta_{\mu_4  \mu_2}
-
 \delta_{\mu_1  \mu_2}  \delta_{\mu_3  \mu_4}
-
 \delta_{\mu_1  \mu_4}  \delta_{\mu_2  \mu_3}
\Big]
\delta_{\sum k}
.
\end{align}

\section{Details on \(N=4\) gluons correlator}
\label{sec:details_4_gluons_correlator}
In this section we do not write the hat explicitely in order to make
the notation lighter, i.e. $\hk$ is simply written as $k$.

The direct computation of the correlator gives the following result
\begin{align}
A_{1 2 3 4}&=
\mymeno\int_{0}^{1} d y~
C_{}
\delta_{k_{1}+k_{2}+k_{3}+k_{4}}
\nonumber\\
C_{}&=
C_{ (0, 0) } 
+ \frac{1}{y} C_{ (1, 0) }+ \frac{1}{y^{2}} C_{ (2, 0) }
+ \frac{1}{1-y} C_{ (0, 1) }+ \frac{1}{(1-y)^{2}} C_{ (0, 2) }
,
\end{align}
after we write the rational expressions involving $y$ as sum of simple
factors, e.g. $y/(1-y)= 1 - 1/(1-y)$.
The the different contributions are given as follows.

Terms proportional to \(y^{-1}\):
\begin{align}
C_{ (1, 0) }=
& 
\mypiu { \epsilon_1}\cdot { k_3}\,{ \epsilon_2}\cdot { k_4}\,{ \epsilon_3}\cdot
 { k_4}\,{ \epsilon_4}\cdot { k_3}
\nonumber\\ &
\mypiu { \epsilon_1}\cdot { k_3}\,
 { \epsilon_2}\cdot { k_3}\,{ \epsilon_3}\cdot { k_4}\,{ \epsilon_4}\cdot
 { k_3}
\nonumber\\ &
\mypiu { \epsilon_1}\cdot { k_2}\,{ \epsilon_2}\cdot { k_3}\,
 { \epsilon_3}\cdot { k_4}\,{ \epsilon_4}\cdot { k_3}
\nonumber\\ &
\mymeno{ \epsilon_1}\cdot
 { k_2}\,{ \epsilon_2}\cdot { k_4}\,{ \epsilon_3}\cdot { k_2}\,
 { \epsilon_4}\cdot { k_3}
\nonumber\\ &
\mymeno{ \epsilon_1}\cdot { k_2}\,{ \epsilon_2}\cdot
 { k_3}\,{ \epsilon_3}\cdot { k_2}\,{ \epsilon_4}\cdot { k_3}
\nonumber\\ &
\mypiu { \epsilon_1}\cdot
 { k_2}\,{ \epsilon_2}\cdot { k_4}\,{ \epsilon_3}\cdot { k_4}\,
 { \epsilon_4}\cdot { k_2}
\nonumber\\ &
\mypiu { \epsilon_1}\cdot { k_2}\,{ \epsilon_2}\cdot
 { k_3}\,{ \epsilon_3}\cdot { k_4}\,{ \epsilon_4}\cdot { k_2}
\nonumber\\ &
%
\mypiu  { \epsilon_1}\cdot { \epsilon_2}\,{ \epsilon_3}\cdot { k_4}\,{ \epsilon_4}\cdot
 { k_2}
\nonumber\\ &
\mymeno { \epsilon_1}\cdot { \epsilon_2}\,{ \epsilon_3}\cdot { k_2}\,{ \epsilon_4}\cdot
 { k_3}
\nonumber\\ &
%
\mypiu { \epsilon_1}\cdot { \epsilon_3}\,{ \epsilon_2}\cdot { k_4}\,
 { \epsilon_4}\cdot { k_3}
\nonumber\\ &
\mypiu { \epsilon_1}\cdot { \epsilon_3}\,{ \epsilon_2}\cdot
 { k_3}\,{ \epsilon_4}\cdot { k_3}
\nonumber\\ &
%
\mymeno{ \epsilon_1}\cdot { \epsilon_4}\,{ \epsilon_2}\cdot
 { k_3}\,{ \epsilon_3}\cdot { k_4}
\nonumber\\ &
\mymeno{ \epsilon_1}\cdot { \epsilon_4}\,{ \epsilon_2}\cdot { k_4}\,
 { \epsilon_3}\cdot { k_4}
\nonumber\\ &
%
\mypiu { \epsilon_1}\cdot { k_2}\,
 { \epsilon_2}\cdot { \epsilon_3}\,{ \epsilon_4}\cdot { k_3}
\nonumber\\ &
%
\mymeno{ \epsilon_1}\cdot { k_2}\,
 { \epsilon_2}\cdot { \epsilon_4}\,{ \epsilon_3}\cdot { k_4}
\nonumber\\ &
%
\mymeno{ \epsilon_1}\cdot
 { k_3}\,{ \epsilon_2}\cdot { k_4}\,{ \epsilon_3}\cdot { \epsilon_4}
\nonumber\\ &
\mymeno { \epsilon_1}\cdot { k_3}\,{ \epsilon_2}\cdot { k_3}\,{ \epsilon_3}\cdot
 { \epsilon_4}
\nonumber\\ &
\mymeno{ \epsilon_1}\cdot { k_2}\,{ \epsilon_2}\cdot { k_3}\,
 { \epsilon_3}\cdot { \epsilon_4}
\end{align}
This can be simplified to\footnote{
We use
${\epsilon_1} \cdot {k_3} \wedge {k_4} \cdot {\epsilon_2}
=
{\epsilon_1} \cdot {k_3} ~ {k_4} \cdot {\epsilon_2}
-
{\epsilon_1} \cdot {k_4} ~ {k_3} \cdot {\epsilon_2}
$.
}
\begin{align}
\label{C10}
C_{ (1, 0) }=
& 
\mymeno{\epsilon_1} \cdot {k_3} \wedge {k_4} \cdot {\epsilon_2}  
\,{ \epsilon_3} \cdot {k_3}\wedge {k_4} \cdot {\epsilon_4}
\nonumber\\
& 
\mymeno{\epsilon_1} \cdot {k_1} \wedge {k_2} \cdot {\epsilon_2}  
\,{ \epsilon_3} \cdot {k_1}\wedge {k_2} \cdot {\epsilon_4}
\nonumber\\
& 
\mymeno{\epsilon_1} \cdot {\epsilon_2}  
\,{ \epsilon_3} \cdot {k_1}\wedge {k_2} \cdot {\epsilon_4}
\nonumber\\
& 
\mymeno{\epsilon_1} \cdot {\epsilon_3}  
\,{ \epsilon_2} \cdot {k_1}\, {\epsilon_4} \cdot {k_{3}}
\nonumber\\
& 
\mypiu {\epsilon_1} \cdot {\epsilon_4}  
\,{ \epsilon_2} \cdot {k_1}\, {\epsilon_3} \cdot {k_{4}}
\nonumber\\
& 
\mypiu {\epsilon_2} \cdot {\epsilon_3}  
\,{ \epsilon_1} \cdot {k_2}\, {\epsilon_4} \cdot {k_{3}}
\nonumber\\
& 
\mymeno{\epsilon_2} \cdot {\epsilon_4}  
\,{ \epsilon_1} \cdot {k_2}\, {\epsilon_3} \cdot {k_{4}}
\nonumber\\
& 
\mymeno{\epsilon_3} \cdot {\epsilon_4}  
\,{ \epsilon_1} \cdot {k_3}\wedge {k_4} \cdot {\epsilon_2}
\nonumber\\
\end{align}

Terms proportional to \(y^{-2}\):
\begin{align}
C_{ (2, 0) }=
&
\mypiu { \epsilon_1}\cdot { k_2}\,{ \epsilon_2}\cdot { k_4}\,{ \epsilon_3}\cdot
 { k_4}\,{ \epsilon_4}\cdot { k_3}
\nonumber\\&\mypiu { \epsilon_1}\cdot { k_2}\,
 { \epsilon_2}\cdot { k_3}\,{ \epsilon_3}\cdot { k_4}\,{ \epsilon_4}\cdot
 { k_3}
\nonumber\\&
\mypiu { \epsilon_1}\cdot { \epsilon_2}\,{ \epsilon_3}\cdot { k_4 }\,
 { \epsilon_4}\cdot { k_3}
\nonumber\\&\mymeno{ \epsilon_1}\cdot { k_2}
\,{ \epsilon_2}\cdot { k_4}\,{ \epsilon_3}\cdot { \epsilon_4}
\nonumber\\&\mymeno{ \epsilon_1}\cdot { k_2}\,
 { \epsilon_2}\cdot { k_3}\,{ \epsilon_3}\cdot { \epsilon_4}
\nonumber\\&\mymeno{ \epsilon_1}\cdot
 { \epsilon_2}\,{ \epsilon_3}\cdot { \epsilon_4}
\end{align}
This can be simplified to
\begin{align}
\label{C20}
C_{ (2, 0) }=
&
\mymeno{ \epsilon_1}\cdot { k_2}\,{ \epsilon_2}\cdot { k_1}
\,{ \epsilon_3}\cdot { k_4}\,{ \epsilon_4}\cdot { k_3}
\nonumber\\
&
\mypiu { \epsilon_1}\cdot { \epsilon_2}\,{ \epsilon_3}\cdot { k_4 }\,
 { \epsilon_4}\cdot { k_3}
\nonumber\\
&
\mypiu { \epsilon_3}\cdot { \epsilon_4}\,
{ \epsilon_1}\cdot { k_2}
\,{ \epsilon_2}\cdot { k_1}
\nonumber\\
&\mymeno{ \epsilon_1}\cdot { \epsilon_2}\,{ \epsilon_3}\cdot { \epsilon_4}
\end{align}

Terms proportional to \((1-y)^{-1}\):
\begin{align}
C_{ (0, 1) }=
&
\mypiu { \epsilon_1}\cdot { k_3}\,{ \epsilon_2}\cdot { k_3}
\,{ \epsilon_3}\cdot { k_4}\,{ \epsilon_4}\cdot { k_3}
\nonumber\\ 
&
\mypiu { \epsilon_1}\cdot { k_2}\,{ \epsilon_2}\cdot { k_3}
\,{ \epsilon_3}\cdot { k_4}\,{ \epsilon_4}\cdot { k_3}
\nonumber\\ 
&
\mymeno{ \epsilon_1}\cdot { k_3}\,{ \epsilon_2}\cdot { k_4}\,
 { \epsilon_3}\cdot { k_2}\,{ \epsilon_4}\cdot { k_3}
\nonumber\\ 
&
\mymeno{ \epsilon_1}\cdot { k_2}\,{ \epsilon_2}\cdot { k_4}\,
{ \epsilon_3}\cdot { k_2}\, { \epsilon_4}\cdot { k_3}
\nonumber\\ &
\mymeno{ \epsilon_1}\cdot { k_2}\,{ \epsilon_2}\cdot{ k_3}
\,{ \epsilon_3}\cdot { k_2}\,{ \epsilon_4}\cdot { k_3}
\nonumber\\ &
\mypiu { \epsilon_1}\cdot { k_3}\,
 { \epsilon_2}\cdot { k_3}\,{ \epsilon_3}\cdot { k_4}\,{ \epsilon_4}\cdot
 { k_2}
\nonumber\\ &
\mypiu { \epsilon_1}\cdot { k_2}\,{ \epsilon_2}\cdot { k_3}\,
 { \epsilon_3}\cdot { k_4}\,{ \epsilon_4}\cdot { k_2}
\nonumber\\ &
\mymeno{ \epsilon_1}\cdot
 { k_3}\,{ \epsilon_2}\cdot { k_4}\,{ \epsilon_3}\cdot { k_2}\,
 { \epsilon_4}\cdot { k_2}
\nonumber\\ &
\mymeno{ \epsilon_1}\cdot { k_2}\,{ \epsilon_2}\cdot
 { k_4}\,{ \epsilon_3}\cdot { k_2}\,{ \epsilon_4}\cdot { k_2}
\nonumber\\ &
\mypiu 
 { \epsilon_1}\cdot { k_3}\,{ \epsilon_2}\cdot { k_3}\,{ \epsilon_3}\cdot
 { k_2}\,{ \epsilon_4}\cdot { k_2}
\nonumber\\ &
%
\mymeno { \epsilon_1}\cdot { \epsilon_2}\,{ \epsilon_3}\cdot { k_2}
\,{ \epsilon_4}\cdot { k_3}
\nonumber\\ &
%
\mymeno{ \epsilon_1}\cdot { \epsilon_2}\,
 { \epsilon_3}\cdot { k_2}\,{ \epsilon_4}\cdot { k_2}
\nonumber\\ &
%
\mypiu { \epsilon_1}\cdot
 { \epsilon_3}\,{ \epsilon_2}\cdot { k_3}\,{ \epsilon_4}\cdot { k_2}
\nonumber\\ &
\mypiu { \epsilon_1}\cdot { \epsilon_3}\,{ \epsilon_2}\cdot { k_3 }\,
 { \epsilon_4}\cdot { k_3}
\nonumber\\ &
%
\mymeno{ \epsilon_1}\cdot { \epsilon_4}\,{ \epsilon_2}\cdot { k_3}\,
 { \epsilon_3}\cdot { k_4}
\nonumber\\ &
\mypiu { \epsilon_1}\cdot { \epsilon_4}\,{ \epsilon_2}\cdot
 { k_4}\,{ \epsilon_3}\cdot { k_2}
\nonumber\\ &
%
\mypiu { \epsilon_1}\cdot { k_2}\,
{ \epsilon_2}\cdot{ \epsilon_3}\,{ \epsilon_4}\cdot { k_3}
\nonumber\\ &
\mymeno{ \epsilon_1}\cdot { k_3}\,{ \epsilon_2}\cdot { \epsilon_3}\,{ \epsilon_4}\cdot
 { k_2}
\nonumber\\ &
%
\mypiu { \epsilon_1}\cdot { k_3}\,
 { \epsilon_2}\cdot { \epsilon_4}\,{ \epsilon_3}\cdot { k_2}
\nonumber\\ &
\mypiu { \epsilon_1}\cdot
 { k_2}\,{ \epsilon_2}\cdot { \epsilon_4}\,{ \epsilon_3}\cdot {k_2}
\nonumber\\ &
%
\mymeno{ \epsilon_1}\cdot { k_3}\,{ \epsilon_2}\cdot { k_3}
\,{ \epsilon_3}\cdot { \epsilon_4}
\nonumber\\ &
\mymeno{ \epsilon_1}\cdot { k_2}\,{ \epsilon_2}\cdot { k_3}\,
 { \epsilon_3}\cdot { \epsilon_4}
\end{align}
This can be simplified to
\begin{align}
\label{C01}
C_{ (0, 1) }=
&
\mypiu { \epsilon_1}\cdot { k_4}
\,{ \epsilon_4}\cdot { k_1}
\,{ \epsilon_2}\cdot [ { k_4} \wedge  { k_1} ] \cdot { \epsilon_3}
\nonumber\\
&\mypiu { \epsilon_2}\cdot { k_3}
\,{ \epsilon_3}\cdot { k_2}
\,{ \epsilon_1}\cdot [ { k_3} \wedge  { k_2} ] \cdot { \epsilon_4}
\nonumber\\
&\mypiu  { \epsilon_1}\cdot { \epsilon_2}
\,{ \epsilon_3}\cdot { k_2} \,{ \epsilon_4}\cdot { k_1}
\nonumber\\
&\mymeno { \epsilon_1}\cdot { \epsilon_3}
\,{ \epsilon_2}\cdot { k_3} \,{ \epsilon_4}\cdot { k_1}
\nonumber\\
&\mypiu  { \epsilon_1}\cdot { \epsilon_4}
\,{ \epsilon_2}\cdot [ { k_1} \wedge {k_4} ] \cdot { \epsilon_3}
\nonumber\\
&\mypiu  { \epsilon_2}\cdot { \epsilon_3}
\,{ \epsilon_1}\cdot [ { k_2} \wedge {k_3} ] \cdot { \epsilon_3}
\nonumber\\
&\mymeno { \epsilon_2}\cdot { \epsilon_4}
\,{ \epsilon_1}\cdot { k_4} \,{ \epsilon_3}\cdot { k_2}
\nonumber\\
&\mypiu  { \epsilon_3}\cdot { \epsilon_4}
\,{ \epsilon_1}\cdot { k_4} \,{ \epsilon_2}\cdot { k_3}
\end{align}

Terms proportional to \((1-y)^{-2}\):
\begin{align}
C_{ (0, 2) }=
&\mymeno{ \epsilon_1}\cdot { k_3}\,{ \epsilon_2}\cdot { k_3}\,{ \epsilon_3}
 \cdot { k_2}\,{ \epsilon_4}\cdot { k_3}
\nonumber\\
&\mymeno{ \epsilon_1}\cdot { k_2}\,
 { \epsilon_2}\cdot { k_3}\,{ \epsilon_3}\cdot { k_2}\,{ \epsilon_4}\cdot
 { k_3}
\nonumber\\
&\mymeno{ \epsilon_1}\cdot { k_3}\,
 { \epsilon_2}\cdot { k_3}\,{ \epsilon_3}\cdot { k_2}\,
{ \epsilon_4}\cdot { k_2}
\nonumber\\
&\mymeno{ \epsilon_1}\cdot { k_2}\,{ \epsilon_2}\cdot { k_3}\,
 { \epsilon_3}\cdot { k_2}\,{ \epsilon_4}\cdot { k_2}
\nonumber\\
&\mypiu { \epsilon_1}\cdot { \epsilon_4}\,{ \epsilon_2}\cdot { k_3}\,
 { \epsilon_3}\cdot { k_2}
\nonumber\\
&\mypiu { \epsilon_1}\cdot{ k_3}\,
{ \epsilon_2}\cdot { \epsilon_3}\,{ \epsilon_4}\cdot { k_2}
\nonumber\\
%
&\mypiu { \epsilon_1}\cdot { k_3}\,{ \epsilon_2}\cdot { \epsilon_3}\,
 { \epsilon_4}\cdot { k_3}
\nonumber\\
%
&\mypiu { \epsilon_1}\cdot { k_2}
\,{ \epsilon_2}\cdot{ \epsilon_3}\,{ \epsilon_4}\cdot { k_3}
\nonumber\\
&\mypiu 
 { \epsilon_1}\cdot { k_2}\,{ \epsilon_2}\cdot { \epsilon_3}\,{ \epsilon_4}\cdot
 { k_2}
\nonumber\\
&\mymeno{ \epsilon_1}\cdot { \epsilon_4}\,{ \epsilon_2}\cdot
 { \epsilon_3}
\end{align}
This can be simplified to
\begin{align}
\label{C02}
C_{ (0, 2) }=
&\mymeno{ \epsilon_1}\cdot { k_4}\,{ \epsilon_2}\cdot { k_3}\,
 { \epsilon_3}\cdot { k_2}\,{ \epsilon_4}\cdot { k_1}
\nonumber\\
&\mypiu { \epsilon_1}\cdot { \epsilon_4}\,{ \epsilon_2}\cdot { k_3}\,
 { \epsilon_3}\cdot { k_2}
\nonumber\\
&\mypiu { \epsilon_2}\cdot { \epsilon_3}\,
{ \epsilon_1}\cdot { k_4}\, { \epsilon_4}\cdot { k_1}
\nonumber\\
&\mymeno{ \epsilon_1}\cdot { \epsilon_4}\,{ \epsilon_2}\cdot
 { \epsilon_3}
\end{align}

Terms proportional to \(1\):
\begin{align}
C_{ (0, 0) }=
&
\mypiu { \epsilon_1}\cdot { k_3}\,{ \epsilon_2}\cdot { k_4}\,{ \epsilon_3}\cdot
 { k_4}\,{ \epsilon_4}\cdot { k_2}
\nonumber\\
&\mypiu { \epsilon_1}\cdot { k_3}\,
 { \epsilon_2}\cdot { k_4}\,{ \epsilon_3}\cdot { k_2}\,{ \epsilon_4}\cdot
 { k_2}
\nonumber\\
&\mypiu { \epsilon_1}\cdot { \epsilon_3}\,{ \epsilon_2}\cdot { k_4}\,
 { \epsilon_4}\cdot { k_2}
\nonumber\\
&\mymeno{ \epsilon_1}\cdot { k_3}\,{ \epsilon_2}\cdot
 { \epsilon_4}\,{ \epsilon_3}\cdot { k_4}
\nonumber\\
&\mymeno{ \epsilon_1}\cdot { k_3}\,
 { \epsilon_2}\cdot { \epsilon_4}\,{ \epsilon_3}\cdot { k_2}
\nonumber\\
&\mymeno{ \epsilon_1}\cdot { \epsilon_3}\,{ \epsilon_2}\cdot { \epsilon_4}
\end{align}
This can be simplified to
\begin{align}
\label{C00}
C_{ (0, 0) }=
&
\mymeno{ \epsilon_1}\cdot { k_3}\,{ \epsilon_2}\cdot { k_4}\,
{ \epsilon_3}\cdot { k_1}\,{ \epsilon_4}\cdot { k_2}
\nonumber\\
&\mypiu { \epsilon_1}\cdot { \epsilon_3}\,
{ \epsilon_2}\cdot { k_4}\, { \epsilon_4}\cdot { k_2}
\nonumber\\
&\mypiu { \epsilon_2}\cdot { \epsilon_4}\,
{ \epsilon_1}\cdot { k_3}\,{ \epsilon_3}\cdot { k_1}
\nonumber\\
&\mymeno{ \epsilon_1}\cdot { \epsilon_3}\,{ \epsilon_2}\cdot { \epsilon_4}
\end{align}

\section{Details on the computation of $V_{[4]}$}
The first step is to compute the \subvertices{} with two on shell legs
\begin{align}
- 
\epsilon^{\mu_{1}}_{\noa_{1}}\epsilon^{\mu_{2}}_{\noa_{2}}
V^{(1 2 3)}
_{\mu_{1};~\mu_{2};~\mu}(k_{1}, k_{2}, \qs)
=
&
\ii \Bigl\{
+[ c_2 \epsilon_1 \cdot \epsilon_2 ] k_{1 \mu}
+[ c_1 \epsilon_1 \cdot \epsilon_2 ] k_{2 \mu}
\nonumber\\
&
+[ (c_1-c_2) \epsilon_2 \cdot k_1 ] \epsilon_{1 \mu}
+[ -(c_1-c_2) \epsilon_1 \cdot k_2 ] \epsilon_{2 \mu}
\Bigr\}
\nonumber\\
+&\ii^3 \Bigl\{
+[ (-3 c_3+2c_5-c_6)  \epsilon_1 \cdot k_2 \epsilon_2 \cdot k_1
+ c_9 \epsilon_1 \cdot \epsilon_2 k_1\cdot k_2 ] k_{1 \mu}
\nonumber\\
&
+[ (-3 c_4-c_5+2 c_6)   \epsilon_1 \cdot k_2 \epsilon_2 \cdot k_1
+ c_{12} \epsilon_1 \cdot \epsilon_2 k_1\cdot k_2 ] k_{1 \mu}
\nonumber\\
&
+[ (-2c_7+c_9+2c_{10}-c_{12}) \epsilon_2 \cdot k_1 k_1\cdot k_2 ] \epsilon_{1 \mu}
\nonumber\\
&
+[ (+2c_8-c_9-2c_{11}+c_{12}) \epsilon_1 \cdot k_2 k_1\cdot k_2 ] \epsilon_{2 \mu}
\Bigr\}
\delta_{\sum k_{i}}
~\qs^\mu
,
\end{align}
and
\begin{align}
- 
V^{(1 2 3)}
_{\mu;~\mu_{3};~\mu_{4}}(-\qs, k_{3}, k_{4})
\epsilon^{\mu_{3}}_{\noa_{3}}\epsilon^{\mu_{4}}_{\noa_{4}}
=
&
\ii \Bigl\{
+[ c_2 \epsilon_3 \cdot \epsilon_4 ] k_{3 \mu}
+[ c_1 \epsilon_3 \cdot \epsilon_4 ] k_{4 \mu}
\nonumber\\
&
+[ (c_1-c_2) \epsilon_4 \cdot k_3 ] \epsilon_{3 \mu}
+[ -(c_1-c_2) \epsilon_3 \cdot k_4 ] \epsilon_{4 \mu}
\Bigr\}
\nonumber\\
+&\ii^3 \Bigl\{
+[ (-3 c_3+2c_5-c_6)  \epsilon_3 \cdot k_4 \epsilon_4 \cdot k_3
+ c_9 \epsilon_3 \cdot \epsilon_4 k_3\cdot k_4 ] k_{3 \mu}
\nonumber\\
&
+[ (-3 c_4-c_5+2 c_6)   \epsilon_3 \cdot k_4 \epsilon_4 \cdot k_3
+ c_{12} \epsilon_3 \cdot \epsilon_4 k_3\cdot k_4 ] k_{1 \mu}
\nonumber\\
&
+[ (-2c_7+c_9+2c_{10}-c_{12}) \epsilon_4 \cdot k_3 k_3\cdot k_4 ] \epsilon_{3 \mu}
\nonumber\\
&
+[ (+2c_8-c_9-2c_{11}+c_{12}) \epsilon_3 \cdot k_4 k_3\cdot k_4 ] \epsilon_{4 \mu}
\Bigr\}
\delta_{\sum k_{i}}
~(-\qs)^\mu.
\end{align}

The computation of the \subvertex{} at $k^0$ order is  
\begin{align}
\ap &\cCz \cNz^4 A_{1 2 3 4}|_{k^0} 
&-&
\epsilon^{\mu_{1}}_{\noa_{1}}\epsilon^{\mu_{2}}_{\noa_{2}}
V^{(1 2 3)}
_{\mu_{1};~\mu_{2};~\mu}(k_{1}, k_{2}, \qs)|_{k^1}
\frac{\delta^{\mu\nu} }{\qs^2}
V^{(1 2 3)}
_{\nu;~\mu_{3};~\mu_{4}}(-\qs, k_{3}, k_{4})|_{k^1}
~\epsilon^{\mu_{3}}_{\noa_{3}}\epsilon^{\mu_{4}}_{\noa_{4}}
\delta_{\sum k_{i}}
\nonumber\\
&
&-&
~\epsilon^{\mu_{4}}_{\noa_{4}}\epsilon^{\mu_{1}}_{\noa_{1}}
V^{(1 2 3)}
_{\mu_{4};~\mu_{1};~\mu}(k_{4}, k_{1}, \qs)|_{k^1}
\frac{\delta^{\mu\nu} }{\qs^2}
V^{(1 2 3)}
_{\nu;~\mu_{2};~\mu_{3}}(-\qs, k_{2}, k_{3})|_{k^1}
 ~\epsilon^{\mu_{2}}_{\noa_{2}}\epsilon^{\mu_{3}}_{\noa_{3}}
\delta_{\sum k_{i}}
,
\end{align}
and it can be split into three pieces.
The first piece contains the pole in the $s$ channel%
\begin{align}
-
\ap &
\cCz \cNz^4 
\frac{1}{s}
\left[
C_{(2,0)}|_{k^0} u + 2 C_{(1,0)}|_{k^2}  
\right]
\delta_{\sum \hk_{i}}
\nonumber\\
-&
\epsilon^{\mu_{1}}_{\noa_{1}}\epsilon^{\mu_{2}}_{\noa_{2}}
V^{(1 2 3)}
_{\mu_{1};~\mu_{2};~\mu}(k_{1}, k_{2}, \qs)|_{k^1}
\frac{\delta^{\mu\nu} }{\qs^2}
V^{(1 2 3)}
_{\nu;~\mu_{3};~\mu_{4}}(-\qs, k_{3}, k_{4})|_{k^1}
~\epsilon^{\mu_{3}}_{\noa_{3}}\epsilon^{\mu_{4}}_{\noa_{4}}
\delta_{\sum k_{i}}
\nonumber\\
=&
-\oh (c_1^2 + c_2^2) 
\epsilon_1\cdot \epsilon_3\, \epsilon_2\cdot \epsilon_4
,
\end{align}
when the constraint (\ref{eq:constraint_N4_k0}) \footnote{
The powers of $\tap$ come from the fact that $\hk=\sqtap k$ and $\he=
\sqtap \epsilon$.
}
\begin{equation}
(c_{1} - c_{2} )^{2} = - \cCz \cNz^{4} \tap^{3-\oh D}
\end{equation}
 is satisfied.
The second piece contains the pole in the $u$ channel and can be
obtained from the first one by a cyclic permutation $(1 2 3 4)$.
Finally the third piece come from the string amplitude without poles,
i.e.
\begin{align}
\ap\cCz \cNz^4  
&A_{ 1 2 3 4 } |_{k^0}
=
\ap\cCz \cNz^4  
\left[
-C_{(2,0)}|_{k^0} -C_{(0,2)}|_{k^0} +C_{(0,0)}|_{k^0}
\right]
\delta_{\sum \hk_{i}}
\nonumber\\
&=
-\frac{\tap^{3-\oh D}}{2} \cCz \cNz^4
\left[
\epsilon_1\cdot \epsilon_2\, \epsilon_3\cdot \epsilon_4
+
\epsilon_4\cdot \epsilon_1\, \epsilon_2\cdot \epsilon_3
-
\epsilon_1\cdot \epsilon_3\, \epsilon_2\cdot \epsilon_4
\right]
\delta_{\sum k_{i}}
.
\end{align}
Assembling all pieces gives the result (\ref{eq:subvertex_N4_k0}).

The computation of the \subvertex{} at $k^2$ order proceeds in the
same way.
The first term is given by the terms with a pole in the $s$ channel

\begin{align}
\ap
&\cCz \cNz^4 
\frac{1}{s}
\left[
C_{(2,0)}|_{k^2} u - 2 C_{(1,0)}|_{k^4}  
\right]
\delta_{\sum \hk_{i}}
\nonumber\\
-&
\Big[
V^{(1 2 3)}
_{\mu_{1};~\mu_{2};~\mu}(k_{1}, k_{2}, \qs)
V^{(1 2 3)}
_{\nu;~\mu_{3};~\mu_{4}}(-\qs, k_{3}, k_{4})
\Big]|_{k^4}
\epsilon^{\mu_{1}}_{\noa_{1}}\epsilon^{\mu_{2}}_{\noa_{2}}
\frac{\delta^{\mu\nu} }{\qs^2}
~\epsilon^{\mu_{3}}_{\noa_{3}}\epsilon^{\mu_{4}}_{\noa_{4}}
\delta_{\sum k_{i}}
\nonumber\\
=
&-\oh [ c_1 ( - 3 c_4- c_5 + 2 c_6) + c_2 ( -3 c_3 + 2 c_5 - c_6) ]
[
\epsilon_1 \cdot \epsilon_2\, \epsilon_3\cdot k_4\, \epsilon_4\cdot  k_3
+
\epsilon_3 \cdot \epsilon_4\, \epsilon_1\cdot k_2\, \epsilon_2\cdot  k_1
]
\nonumber\\
&
\end{align}
when the constraint 
\begin{align}
(c_1 - c_2) ( -3 c_3 + 3 c_4 + 3 c_5 -  3 c_6 )
=
\cCz \cNz^4
\tap^{4-\oh D} 
\end{align}
is satisfied.
The second piece contains the pole in the $u$ channel and can be
obtained from the first one by a cyclic permutation $(1 2 3 4)$.
The third piece come from the string amplitude without poles
and  from the Feynman diagram obtained from \subvertices{} without poles.
Explicitly we have that  the terms of the form
$(\epsilon\cdot\epsilon)^2 k\cdot k $ are
\begin{align}
%
%
\bigg\{
&
+\oh \epsilon_1 \cdot \epsilon_2\, \epsilon_3\epsilon_4
\Big[ 
\left( \oh \tap^{4-\oh D} \cCz \cNz^4 \right) s
+\left( \oh \tap^{4-\oh D} \cCz \cNz^4  +c_1 c_9+ c_2 c_{12}\right) u
\nonumber\\
&\phantom{+ +\oh \epsilon_1 \cdot \epsilon_2\, \epsilon_3\epsilon_4\Big[ }
+\left( \oh \tap^{4-\oh D} \cCz \cNz^4  +c_1 c_{12}+ c_2 c_{9}\right) t
\Big] 
\nonumber\\
&+
\mbox{1 term obtained by cycling $(1234)$ in the previous term}
\bigg\}
\nonumber\\
&
-\oh \epsilon_3 \cdot \epsilon_1\, \epsilon_2\epsilon_4
\left[ 
\left( \oh \tap^{4-\oh D} \cCz \cNz^4 \right) u
+\left( \oh \tap^{4-\oh D} \cCz \cNz^4 \right) s
\right]
\end{align}

The terms of the form $(\epsilon\cdot\epsilon) (\epsilon\cdot k)^2 $ 
in a canonical form where the indeces of the momenta are taken from
the indeces of $(\epsilon\cdot\epsilon)$  are
\begin{align}
+\bigg\{
+&\epsilon_1 \cdot \epsilon_2\, \epsilon_3\cdot k_1\, \epsilon_4\cdot  k_1
\oh
\Big[
 \tap^{4-\oh D} \cCz \cNz^4
+ c_1 ( -3 c_4 -c_5+2 c_6)
\nonumber\\
&\phantom{\epsilon_1 \cdot \epsilon_2\, \epsilon_3\cdot k_1\,
  \epsilon_4\cdot  k_1 \oh } 
+ c_2 ( -3 c_3 + 2c_5-c_6-2c_7+2c_8+2c_{10}-2c_{11})
\Big]
\nonumber\\
+&\epsilon_1 \cdot \epsilon_2\, \epsilon_3\cdot k_1\, \epsilon_4\cdot  k_2
\oh
\Big[
 \tap^{4-\oh D} \cCz \cNz^4
+ c_1 ( -3c_4 - c_5 +2c_6 +2c_8 -2c_{11})
\nonumber\\
&\phantom{\epsilon_1 \cdot \epsilon_2\, \epsilon_3\cdot k_1\,
  \epsilon_4\cdot  k_1 \oh } 
+ c_2 ( -3c_3 +2c_5 -c_6 -2c_7 +2c_{10})
\Big]
\nonumber\\
+&\epsilon_1 \cdot \epsilon_2\, \epsilon_3\cdot k_1\, \epsilon_4\cdot  k_2
\oh
\Big[
 \tap^{4-\oh D} \cCz \cNz^4
+ c_1 ( -3c_4 - c_5 +2c_6 -4c_7 -2c_8 +4c_9 +2c_{10} +2c_{11} -2c_{12})
\nonumber\\
&\phantom{\epsilon_1 \cdot \epsilon_2\, \epsilon_3\cdot k_1\,
  \epsilon_4\cdot  k_1 \oh } 
+ c_2 ( -3c_3 +2c_5 -c_6 +2c_7 +4c_8 -2c_9 -2c_{10} -4c_{11} +4c_{12})
\Big]
\nonumber\\
+&\epsilon_1 \cdot \epsilon_2\, \epsilon_3\cdot k_2\, \epsilon_4\cdot  k_2
\oh
\Big[
 \tap^{4-\oh D} \cCz \cNz^4
+ c_1 ( -3c_4 - c_5 +2c_6 -2c_7 +2c_8 )
\nonumber\\
&\phantom{\epsilon_1 \cdot \epsilon_2\, \epsilon_3\cdot k_1\,
  \epsilon_4\cdot  k_1 \oh } 
+ c_2 ( -3c_3 +2c_5 -c_6 )
\Big]
\nonumber\\
&+
\mbox{3*4 terms obtained by cycling $(1234)$ in the previous 4 terms}
\bigg\}
\nonumber\\
+\bigg\{
+&\epsilon_1 \cdot \epsilon_3\, \epsilon_2\cdot k_1\, \epsilon_4\cdot  k_1
\Big [-\oh  \tap^{4-\oh D} \cCz \cNz^4 \Big ]
\nonumber\\
+&\epsilon_1 \cdot \epsilon_3\, \epsilon_2\cdot k_1\, \epsilon_4\cdot  k_3
\Big [-\oh  \tap^{4-\oh D} \cCz \cNz^4 
-(c_1-c_2)( -2c_7 +c_9 +2c_{10} -c_{12}) \Big ]
\nonumber\\
+&\epsilon_1 \cdot \epsilon_3\, \epsilon_2\cdot k_3\, \epsilon_4\cdot  k_1
\Big [-\oh  \tap^{4-\oh D} \cCz \cNz^4 
+(c_1-c_2)( +2c_7 -c_9 -2c_{11} +c_{12}) ]
\Big ]
\nonumber\\
+&\epsilon_1 \cdot \epsilon_3\, \epsilon_2\cdot k_3\, \epsilon_4\cdot  k_3
\Big [-\oh  \tap^{4-\oh D} \cCz \cNz^4 \Big ]
\nonumber\\
&+
\mbox{4 terms obtained by cycling $(1234)$ in the previous 4 terms}
\bigg\}
\end{align}


\begin{thebibliography}{99}
\bibitem{Scherk:1971xy}
  J.~Scherk,
  Nucl.\ Phys.\ B {\bf 31} (1971) 222.
  doi:10.1016/0550-3213(71)90227-6

\bibitem{Neveu:1971mu}
  A.~Neveu and J.~Scherk,
  Nucl.\ Phys.\ B {\bf 36} (1972) 155.
  doi:10.1016/0550-3213(72)90301-X

\bibitem{Scherk:1974ca}
  J.~Scherk and J.~H.~Schwarz,
  Nucl.\ Phys.\ B {\bf 81} (1974) 118.
  doi:10.1016/0550-3213(74)90010-8

\bibitem{Scherk:1974mc}
  J.~Scherk and J.~H.~Schwarz,
  Phys.\ Lett.\  {\bf 52B} (1974) 347.
  doi:10.1016/0370-2693(74)90059-8

\bibitem{Yoneya:1974jg}
  T.~Yoneya,
  Prog.\ Theor.\ Phys.\  {\bf 51} (1974) 1907.
  doi:10.1143/PTP.51.1907


\bibitem{Tseytlin:1986ti}
  A.~A.~Tseytlin,
  Nucl.\ Phys.\ B {\bf 276} (1986) 391
   Erratum: [Nucl.\ Phys.\ B {\bf 291} (1987) 876].
  doi:10.1016/0550-3213(86)90303-2, 10.1016/0550-3213(87)90500-1

\bibitem{Coletti:2003ai}
  E.~Coletti, I.~Sigalov and W.~Taylor,
  JHEP {\bf 0309} (2003) 050
  doi:10.1088/1126-6708/2003/09/050
  [hep-th/0306041].


\bibitem{Gervais:1972tr}
  J.~L.~Gervais and A.~Neveu,
  Nucl.\ Phys.\ B {\bf 46} (1972) 381.
  doi:10.1016/0550-3213(72)90071-5

\bibitem{Magnea:2015fsa}
  L.~Magnea, S.~Playle, R.~Russo and S.~Sciuto,
  JHEP {\bf 1506} (2015) 146
  doi:10.1007/JHEP06(2015)146
  [arXiv:1503.05182 [hep-th]].
\\
  L.~Magnea, S.~Playle, R.~Russo and S.~Sciuto,
  JHEP {\bf 1309} (2013) 081
  doi:10.1007/JHEP09(2013)081
  [arXiv:1305.6631 [hep-th]].
\\
  L.~Magnea, R.~Russo and S.~Sciuto,
  Int.\ J.\ Mod.\ Phys.\ A {\bf 21} (2006) 533
  doi:10.1142/S0217751X06025110
  [hep-th/0412087].

\bibitem{Bern:1991an}
  Z.~Bern and D.~C.~Dunbar,
  Nucl.\ Phys.\ B {\bf 379} (1992) 562.
  doi:10.1016/0550-3213(92)90135-X

\bibitem{Mangano:1990by}
  M.~L.~Mangano and S.~J.~Parke,
  Phys.\ Rept.\  {\bf 200} (1991) 301
  doi:10.1016/0370-1573(91)90091-Y
  [hep-th/0509223].


\bibitem{Dixon:1996wi}
  L.~J.~Dixon,
  In *Boulder 1995, QCD and beyond* 539-582
  [hep-ph/9601359].



\bibitem{Cuomo:2000de}
  F.~Cuomo, R.~Marotta, F.~Nicodemi, R.~Pettorino, F.~Pezzella and G.~Sabella,
  Mod.\ Phys.\ Lett.\ A {\bf 16} (2001) 1035
  doi:10.1142/S0217732301004212
  [hep-th/0011071].
\\
  R.~Marotta and F.~Pezzella,
  Phys.\ Rev.\ D {\bf 61} (2000) 106006
  doi:10.1103/PhysRevD.61.106006
  [hep-th/9912158].
\\
  R.~Marotta and F.~Pezzella,
  PoS tmr {\bf 99} (1999) 033
  [hep-th/0003044].
\\
  A.~Liccardo, F.~Pezzella and R.~Marotta,
  Mod.\ Phys.\ Lett.\ A {\bf 14} (1999) 799
  doi:10.1142/S0217732399000845
  [hep-th/9903027].
\\
  L.~Cappiello, A.~Liccardo, R.~Pettorino, F.~Pezzella and R.~Marotta,
  Lect.\ Notes Phys.\  {\bf 525} (1999) 466
  doi:10.1007/BFb0104267
  [hep-th/9812152].
\\
  L.~Cappiello, R.~Marotta, R.~Pettorino and F.~Pezzella,
  Mod.\ Phys.\ Lett.\ A {\bf 13} (1998) 2845
  doi:10.1142/S0217732398003028
  [hep-th/9808164].
\\
  L.~Cappiello, R.~Marotta, R.~Pettorino and F.~Pezzella,
  Mod.\ Phys.\ Lett.\ A {\bf 13} (1998) 2433
  doi:10.1142/S021773239800259X
  [hep-th/9804032].



\bibitem{SDS}
S.~Sciuto,
Lett.\ Nuovo Cim.\  {\bf 2} (1969) 411.
\\
A. Della Selva and S. Saito,
Lett.\ Nuovo Cim.\  {\bf 4} (1970) 689.

\bibitem{DiVecchia:1996uq}
  P.~Di Vecchia, L.~Magnea, A.~Lerda, R.~Russo and R.~Marotta,
  Nucl.\ Phys.\ B {\bf 469} (1996) 235
  doi:10.1016/0550-3213(96)00141-1
  [hep-th/9601143].



\bibitem{Caneschi:1970sw}
  L.~Caneschi and A.~Schwimmer,
  Lett.\ Nuovo Cim.\  {\bf 3S1} (1970) 213
   [Lett.\ Nuovo Cim.\  {\bf 3} (1970) 213].
  doi:10.1007/BF02755850
\\
  L.~Caneschi, A.~Schwimmer and G.~Veneziano,
  Phys.\ Lett.\  {\bf 30B} (1969) 351.
  doi:10.1016/0370-2693(69)90503-6

\bibitem{GSW}
  M.~B.~Green, J.~H.~Schwarz and E.~Witten,
  ``Superstring Theory. Vol. 1: Introduction,''

\bibitem{Alessandrini:1971fx}
  V.~Alessandrini, D.~Amati, M.~Le Bellac and D.~Olive,
  ``The operator approach to dual multiparticle theory,''
  Phys.\ Rept.\  {\bf 1C} (1971) 269.
  doi:10.1016/0370-1573(71)90008-1









\bibitem{Pesando:2014owa}
  I.~Pesando,
  Nucl.\ Phys.\ B {\bf 866} (2013) 87
  [arXiv:1206.1431 [hep-th]].
\\
  I.~Pesando,
  Nucl.\ Phys.\ B {\bf 886} (2014) 243
  [arXiv:1401.6797 [hep-th]].
\\
  I.~Pesando,
  Nucl.\ Phys.\ B {\bf 889} (2014) 120
  [arXiv:1407.4627 [hep-th]].
\\
  I.~Pesando,
  Int.\ J.\ Mod.\ Phys.\ A {\bf 30} (2015) 21,  1550121
  [arXiv:1107.5525 [hep-th]].


\bibitem{miei_prima}
  I.~Pesando,
  JHEP {\bf 1002} (2010) 064
  [arXiv:0910.2576 [hep-th]].
\\
  I.~Pesando,
  JHEP {\bf 1106} (2011) 138
  [arXiv:1101.5898 [hep-th]].
\\
  P.~Di Vecchia, A.~Liccardo, R.~Marotta, I.~Pesando and F.~Pezzella,
  JHEP {\bf 0711} (2007) 100
  [arXiv:0709.4149 [hep-th]].
\\
  I.~Pesando,
  Phys.\ Lett.\ B {\bf 668} (2008) 324
  [arXiv:0804.3931 [hep-th]].
\\
  I.~Pesando,
  Nucl.\ Phys.\ B {\bf 793} (2008) 211
  [hep-th/0310027].
\\
  P.~Di Vecchia, R.~Marotta, I.~Pesando and F.~Pezzella,
  J.\ Phys.\ A {\bf 44} (2011) 245401
  [arXiv:1101.0120 [hep-th]].
\\
  I.~Pesando,
  Nucl.\ Phys.\ B {\bf 876} (2013) 1
  [arXiv:1305.2710 [hep-th]].




\end{thebibliography}
\end{document}